%% file: main.tex
\documentclass{article}

\usepackage{microtype}
\usepackage{graphicx}
\usepackage{subcaption}
\usepackage{booktabs} %
\usepackage{array}

\usepackage{hyperref}
\usepackage[dvipsnames]{xcolor}

\usepackage[accepted]{icml2024}

\usepackage{amsmath}
\usepackage{amssymb}
\usepackage{mathtools}
\usepackage{amsthm}
\usepackage{multirow}
\usepackage{pifont}%
\usepackage{tabularx}
\usepackage{adjustbox}
\usepackage{epstopdf}
\usepackage{pdfpages}
\newcommand{\cmark}{\ding{51}}%
\newcommand{\xmark}{\ding{55}}%

\usepackage[capitalize,noabbrev]{cleveref}

\theoremstyle{plain}

\theoremstyle{definition}

\theoremstyle{remark}

\input{math_commands.tex}

\icmltitlerunning{PID: Prompt-Independent Data Protection Against Latent Diffusion Models}

\begin{document}

\twocolumn[
\icmltitle{PID: Prompt-Independent Data Protection Against Latent Diffusion Models}

\begin{icmlauthorlist}
\icmlauthor{Ang Li}{sch}
\icmlauthor{Yichuan Mo}{yyy}
\icmlauthor{Mingjie Li}{mj}
\icmlauthor{Yisen Wang}{yyy,ai}
\end{icmlauthorlist}

\icmlaffiliation{yyy}{National Key Lab of General Artificial Intelligence, School of Intelligence Science and Technology, Peking University, China}
\icmlaffiliation{ai}{Institute for Artificial Intelligence, Peking University, China}
\icmlaffiliation{mj}{CISPA Helmholtz Center for Information Security, Germany}
\icmlaffiliation{sch}{School of EECS, Peking University, China}
\icmlcorrespondingauthor{Yisen Wang}{yisen.wang@pku.edu.cn}

\icmlkeywords{Machine Learning, ICML}

\vskip 0.3in
]

\printAffiliationsAndNotice{}  %

\begin{abstract}
     The few-shot fine-tuning of Latent Diffusion Models (LDMs) has enabled them to grasp new concepts from a limited number of images. However, given the vast amount of personal images accessible online, this capability raises critical concerns about civil privacy. While several previous defense methods have been developed to prevent such misuse of LDMs, they typically assume that the textual prompts used by data protectors exactly match those employed by data exploiters. In this paper, we first empirically demonstrate that breaking this assumption, i.e., in cases where discrepancies exist between the textual conditions used by protectors and exploiters, could substantially reduce the effectiveness of these defenses. Furthermore, considering the visual encoder's independence from textual prompts, we delve into the visual encoder and thoroughly investigate how manipulating the visual encoder affects the few-shot fine-tuning process of LDMs. Drawing on these insights, we propose a simple yet effective method called \textbf{Prompt-Independent Defense (PID)} to safeguard privacy against LDMs. We show that PID can act as a strong privacy shield on its own while requiring significantly less computational power. We believe our studies, along with the comprehensive understanding and new defense method, provide a notable advance toward reliable data protection against LDMs. Our code is available at \href{https://github.com/PKU-ML/Diffusion-PID-Protection}{https://github.com/PKU-ML/Diffusion-PID-Protection}

\end{abstract}
\section{Introduction}
\par The advent of Latent Diffusion Models (LDMs) has ushered in an era where images of unprecedented quality are synthesized, blurring the lines between artificial creations and authentic human-generated content, including portraits, photographic arts, and drawings \cite{song2019generative, song2020score, stablediffusion, dalle, midjourney, podell2023sdxl}. A particularly intriguing aspect of LDMs is the capability of few-shot fine-tuning, a.k.a. personalization of the generative model, which teaches the models a brand new concept, such as human faces or painting styles, with as few as 4$\sim$5 images in a matter of minutes \citep{lora, dreambooth, textual_inversion, clark2023directlyfinetune}. However, the ease with which targeted sets of images can be curated with either manual downloading or web crawling on social media renders this capability a double-edged sword. Several selfies casually posted online could mean an array of counterfeit images produced by the LDM fine-tuned by malicious users with the photos, showing exactly the same person clothless or in places he/she has never been to. Civilians are concerned by lawsuits and news related to the unregulated exploitation of such techniques \citep{juefei2022countering}. Thus, developing reliable data protection algorithms that prevent the malicious misuse of LDMs on unauthorized images is vital for both the research community and society. 

\begin{figure*}[th]
    \begin{subfigure}[b]{0.51\textwidth}
        \includegraphics[width=1.0\textwidth]{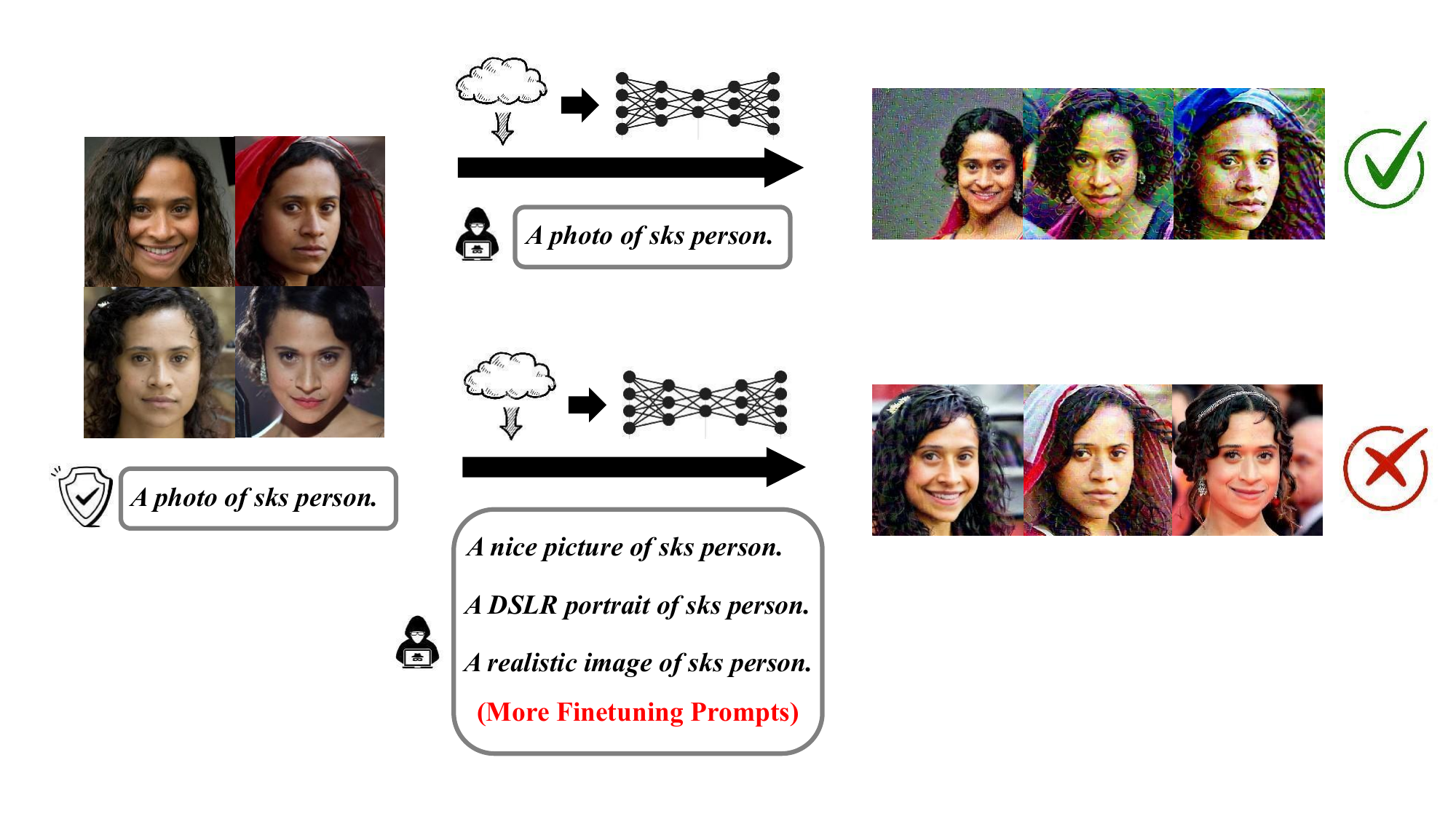}
        \caption{Prompt-dependent defense.}
        \label{figure1-1}
    \end{subfigure}%
    ~
    \hspace{-3mm}
    \begin{subfigure}[b]{0.51\textwidth}
        \centering
        \includegraphics[width=1.0\textwidth]{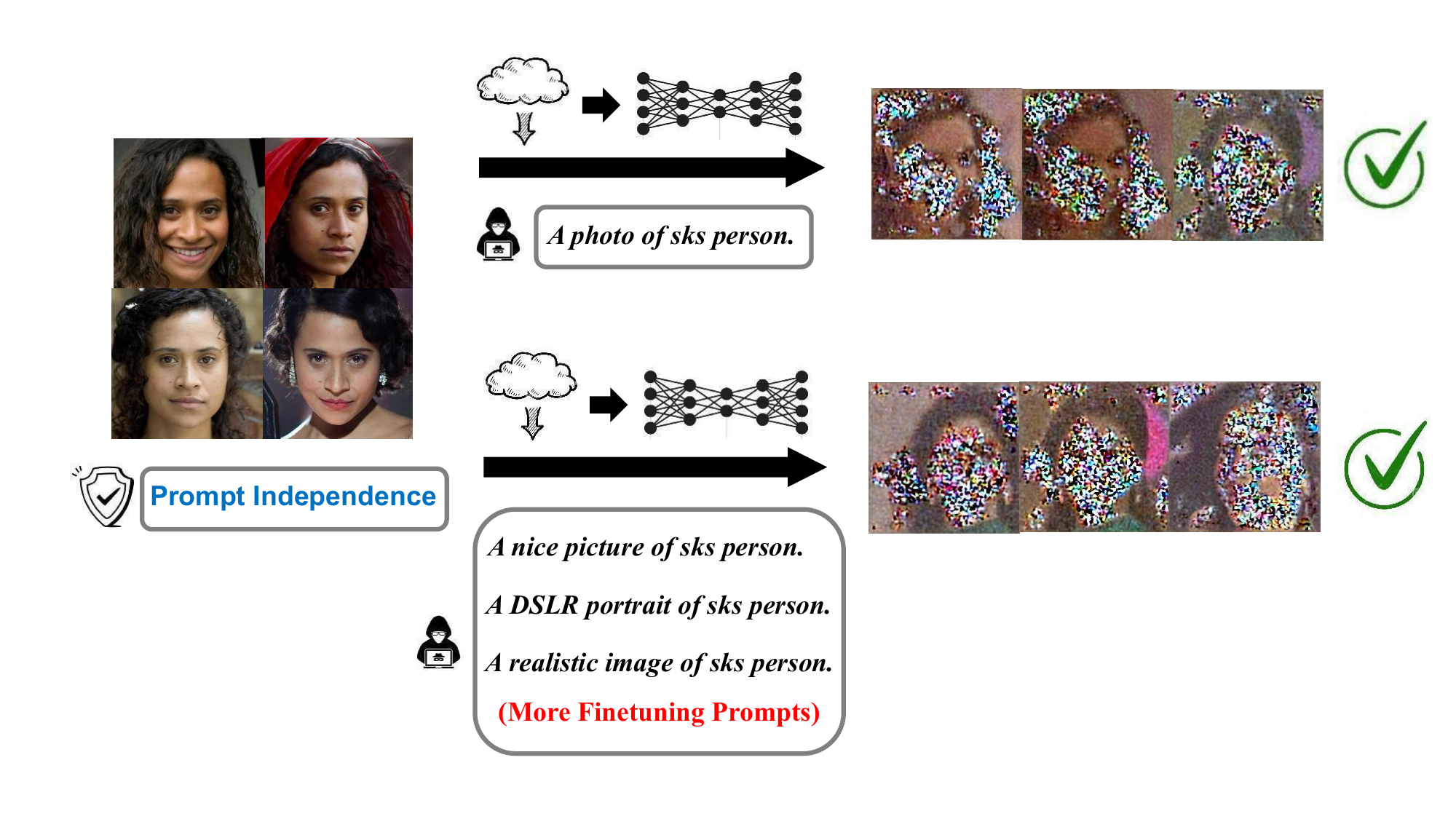}
        \caption{Proposed prompt-independent defense.}
        \label{figure1-2}
    \end{subfigure}%
    \caption{
    Influence of the prompt misalignment, i.e., $\bm{c_{prot}}\neq \bm{c_{explo}}$, on the performance of the prompt-related protection (Figure \ref{figure1-1}) and the prompt-independent one (Figure \ref{figure1-2}).
    In each sub-figure, {the left-most component} depicts the data protection stage and whether a textual prompt is involved. {The middle component} exhibits the data exploiters collect the protected images and try to fine-tune a latent diffusion model with matched/mismatched prompts. {The right-most component} displays some generated images by the generative models fine-tuned with different prompts. The images are all generated with \textit{A high-quality portrait of sks person.} The instance is from the CelebA-HQ dataset \cite{celeba} and the fine-tuned model is Stable Diffusion v1.5 \cite{stablediffusion}.}
    \label{figure1}
\end{figure*}

\par Fortunately, notable efforts have been made to protect images containing sensitive information like human faces or unique art styles against such exploitations by generative models \cite{disruptdeepfake, ruiz2020disrupting, huang2021unlearnable,against_facial_manipulate, wang2022antiforgery, yang2021defendingGAN, li2023unganable}. \citet{photoguard} protect images from malicious image editing. \citet{advDM} adopt the adversarial attack \cite{goodfellow2014explaining,wang2019improving,wang2022unified,mo2022adversarial,Ang} against LDMs to hinder the models from learning features from the protected data. \citet{antidreambooth} propose to generate protective perturbations by attacking a fully trained surrogate model or by synchronizedly disrupting the training process. More recently, \citet{SDS_defense} propose Score Distillation Sampling (SDS) to lighten the computational overhead required by optimizing the protective perturbations. These works have made significant steps toward the ultimate goal. However, when it comes to few-shot fine-tuning, previous works assume to a large degree that the data protection stage (where we add protective perturbations to the images) and the data exploitation stage (where malicious fine-tuning happens), are conditioned on the \textit{identical} textual prompts. Since the data protectors have no prior knowledge of the exploiters, the assumption on prompt consistency may not be realistic in practice.

\par The visual encoder in LDMs projects high-resolution images into a condensed latent space where the diffusion process takes place. Despite its important role, the component has been an \textit{overlooked} part of the protection. Previous studies make arbitrary choices for the visual encoder, like manipulating the mean value of the latent representations \cite{liang2023mist, photoguard}. Note that the independence of the visual encoder from the textual prompts allows it to be unaffected by the assumption of prompt consistency, intuitively making it the right fit for strengthening the protections against varied prompts. Therefore, we raise the following Research Questions (\textbf{RQs}).
\begin{itemize}
\item \textbf{RQ1:} Does a mismatch between the prompts used in the protection and exploitation stages affect the efficacy of existing defense algorithms?
\item \textbf{RQ2:} How do perturbations in pixel space affect the output of the visual encoders in LDMs and thus affect the fine-tuning process? 
\item \textbf{RQ3:} If the answer to RQ1 is \textbf{yes}, can we improve the robustness of the protection by making better use of the prompt-independent visual encoders?
\end{itemize}

\par In this paper, we first investigate the robustness of current defense approaches under the prompt-mismatch scenario. To simulate an adversarial environment where the exploiters intentionally craft textual prompts to undermine the defense, we define a set of candidate prompts, denoted as $\bm{c_{prot}}$, for the exploiters to choose when fine-tuning the latent diffusion models (the full list can be found in Appendix \ref{Appendix A}). We randomly draw an individual from the CelebA-HQ \cite{celeba} dataset and protect the images of it with a typical algorithm ASPL \cite{antidreambooth} using its recommended hyper-parameters. During the protection stage, we fix the textual prompt to be \textit{a photo of sks person} (denoted as $\bm{c_{prot}}$). We then separately fine-tune the Stable Diffusion v1.5 with DreamBooth \cite{dreambooth} conditioned on each of the malicious candidate prompts ($\bm{c_{explo}}$). Lastly, we generate images using the fine-tuned models with the prompt \textit{a high-quality portrait of sks person} and show some of the generated images in Figure \ref{figure1-1}. For the case where $\bm{c_{prot}}\neq \bm{c_{explo}}$, the displayed images are drawn from the visually optimal model among the candidate models. We observe the protective performance of the prompt-dependent defense is notably weakened by the intentionally varied prompts. We hypothesize that the degradation is caused by the entanglement between the perturbations and the textual condition. Deeply concerned by the above observations, we delve into the latent space in LDMs and fully investigate the possibility of utilizing the visual encoder to construct data protection that is more robust to varied prompts. Based upon our findings, we propose a new defense family featuring \textbf{Prompt-Independent Defense (PID)}. PID is completely independent of textual prompts, showing robustness to varied fine-tuning prompts as we show qualitatively in Figure \ref{figure1-2}, and quantitively in Section \ref{Sec6}. Our main contributions are summarized as follows:

\begin{itemize}
    \item We empirically observe that mismatched prompts between the protection stage and the exploitation stage could undermine the effectiveness of current data protection algorithms.
    \item We thoroughly explore the possibility of leveraging the visual encoder within LDMs for more robust data protection and propose a new algorithm named PID.
    \item Through extensive validation, we demonstrate the efficacy of PID against different training algorithms, datasets, and adaptive attacks.
\end{itemize}

\section{Related Work}
\subsection{Diffusion Models}
\par \textbf{Diffusion models } are a type of generative models \cite{sohl2015diffusion, ddpm} that learns the data distribution via two opposing procedures: a forward pass and a backward pass. Given an input image $\bm{x_0} \sim q(x)$, the forward pass gradually adds noise to the image following a noise scheduler $\{\beta_t: \beta_t \in (0, 1)\}_{t=1}^T$ until the data approximately becomes Gaussian noise. For each timestep $t$, the perturbed image is given by $\bm{x_t} = \sqrt{\Tilde{\alpha_t}} \bm{x_0} + \sqrt{1-\Tilde{\alpha_t}}\bm{\epsilon}$,
where $\alpha_t = 1 - \beta_t, \Tilde{\alpha_t} = \Pi_{s=1}^t\alpha_s $ and $\bm{\varepsilon} \sim \gN(0, \bI)$. The reverse process is to reconstruct $\bm{x_0}$ from $\bm{x_T}$ via step-by-step predicting the noise added. Specifically, the noise $\bm{\epsilon}$ at timestep $t$ is estimated by a parameterized network $\bm{\epsilon_\theta}(\bm{x_{t+1}}, t)$. The training loss is commonly defined as the $\ell_2$ distance between the actual noise and the prediction
\begin{equation}
    L_{unc}(\theta, \bm{x_0}) = \E_{\bm{x_0}, t, \bm{\varepsilon}\in\gN(0,\bI)} || \bm{\epsilon} - \bm{\epsilon_\theta}(\bm{x_{t+1}}, t) ||_2^2,
\end{equation}
where $t$ is uniformly sampled from $\{1, 2, \cdots, T\}$ and \textit{unc} stands for unconditional diffusion model. 

\par \textbf{Text-to-Image Latent Diffusion Models} get rid of the massive computational cost brought by operations in pixel space via projecting images into the latent space defined by a pre-trained image encoder \cite{clip, vae}, of which the most widely used implementation is the KL-based VAE \cite{KL-divengence} as adopted by \cite{stablediffusion, podell2023sdxl, DIT}. In this work, we primarily focus on the KL-based VAE while we note that the idea of using the visual encoder for prompt-independent defense is not restricted by the concrete implementation. Denoting the KL-based VAE as $\gE$, the latent distribution of data $\bm{x}$ is given by $\gN(\mu_\gE(\bm{x}), \sigma_\gE^2(\bm{x})) := \gE(\bm{x})$,
and the latent representation of data $\bm{x}$ is sampled from the distribution via re-parametrization, $\bm{z} = \mu_\gE(\bm{x}) + \sigma_\gE(\bm{x})\bm{\varepsilon} := \gE(\bm{x}, \bm{\varepsilon})$,
where $\bm{\varepsilon} \in \gN(0, \bI)$. 
Furthermore, the textual condition $c$ is involved in utilizing the cross-attention \cite{vaswani2017attention, balaji2022ediffi,nichol2022glide, stablediffusion, saharia2022photorealistic} between the UNet \cite{unet} and an extra text encoder \cite{clip}. With the condition $c$ and the latent representation $\bm{z_0} =\gE(\bm{x_0}, \bm{\varepsilon})$, the training process is re-formulated as follows
\begin{equation}
    L_{cond}(\theta, c, \bm{z_0}) = \E_{\bm{z_0}, t, \bm{\varepsilon}} || \bm{\varepsilon} - \bm{\epsilon_\theta}(\bm{z_{t+1}}, c, t) ||_2^2.
\end{equation} 
\par \textbf{Personalization of LDMs} \cite{lora, textual_inversion, dreambooth, clark2023directlyfinetune} enables users to fine-tune the LDMs with only a handful of images. After fine-tuning, the LDMs usually exhibit an astonishing grasp of the concepts contained in the images and can flexibly combine the new concepts with the original training data, synthesizing images that never existed before. The technique seems to be a double-edged sword that raises the potential threat to civil privacy and artists' copyrights to a degree that cannot be ignored anymore.

\subsection{Data Protection against LDM}
\par Worried by the malicious use of LDMs, a series of works have made significant contributions to defend personal images against LDMs. There are two main threads of current works: 1) generating adversarial examples with a surrogate model, specifically AdvDM \cite{advDM}, Mist \cite{liang2023mist}, photoguard \cite{photoguard}, and FSGM \cite{antidreambooth}; and 2) generating unlearnable examples \cite{huang2021unlearnable, TUE} with a bilevel optimization (ASPL \cite{antidreambooth}). Concretely, the former type of defense first fine-tunes a surrogate model $\theta_{sur}$ with the clean data. Then it adversarially maximizes the training loss of $\theta_{sur}$ on the perturbed data:
\begin{align}
    \label{FSGM-Defense}
    \theta_{sur} &= \underset{\theta}{\arg\min} ~L_{cond}(\theta, c, \gE(\bm{x})), \\
    \bm{x^*} &= \underset{||\bm{x^\prime} -\bm{x}||_p \leq \bm{\varepsilon}}{\arg\max}
    L_{cond}(\theta_{sur}, c, \gE(\bm{x^\prime})),
\end{align}
where $\bm{x^*}$ denotes the adversarial examples, i.e., the protected images, and $\bm{\varepsilon}$ ensures the invisibility of the perturbation. Similar to the idea of classic Unlearnable Examples \cite{huang2021unlearnable}, the latter form of defense proposes to generate the protected images alongside the training procedure with a min-max optimization:
\begin{align}
    \label{ASPL-Defense}
    \bm{x^*} &= \underset{ ||\bm{x^\prime} -\bm{x}||_p \leq \bm{\varepsilon}}{\arg\max} \underset{\theta}{\arg\min} 
    L_{cond}(\theta, c, \gE(\bm{x^\prime})).
\end{align}
It is important to note that both of the above algorithms require \textbf{a textual prompt $c$} to protect the images, which makes the perturbations inherently correlated with the text condition. Besides, back-propagating through the large UNet costs enormous GPU VRAM (around 24GB without extra tricks). There also exist protection algorithms involving the visual encoders, either by manipulating the mean value of the latent distribution targetedly \cite{liang2023mist, photoguard}, or by directly making the latent representations unrelated to the data \cite{advDM}. However, previous works have not fully explored the potential of visual encoders, and our empirical studies in Section \ref{Sec:Latent-Space} render the above-mentioned choices suboptimal.

\section{Is Prompt-related Defense Robust to Varied Prompts?}
\label{Sec3}
\par In this section, we conduct quantitive evaluations on the robustness of prompt-related defense when confronted with the changed prompts in a realistic setting.

\par We begin by introducing our overall experimental setup while details are listed in Appendix \ref{Appendix A} for brevity.

\par \textbf{Data \& Model: }Our experiments primarily utilize the CelebA-HQ \cite{celeba} dataset where we randomly select 10 celebrities and choose 4 images for each. Following \citet{antidreambooth}, we use Stable Diffusion v1.5 \cite{stablediffusion} as the default model and DreamBooth \cite{dreambooth} as the default fine-tuning method. 

\par \textbf{Defense: }We consider the method of FSGM and ASPL from \citet{antidreambooth}, whose objectives are entirely correlated with the textual prompts. 
The perturbation budget is set to $0.05$ and the perturbed images are saved in PNG format in this paper unless otherwise specified. 

\par \textbf{Metrics: } We use two metrics to measure the similarity between the generated images and the training images: Face Detection Score (FDS) \cite{mtcnn} and Fréchet Inception Distance (FID) \cite{FID}. Additionally, we use two metrics to assess image quality: Image Quality Score (IQS) \cite{clip} and Blind/Referenceless Image Spatial Quality Evaluator (BRISQUE) \cite{brisque}. We define $\uparrow$ (value increasing) and $\downarrow$ (value decreasing) to indicate the direction of {better protection effect}, e.g., a larger FID indicates a greater distance between the distribution of the generated images and the training images, suggesting that the generated images do not capture the training data well, thus protecting the privacy of the training data.

\begin{figure}[t]
\centering
    \begin{subfigure}[t]{0.45\textwidth}
    \centering
    \includegraphics[width=1.0\textwidth]{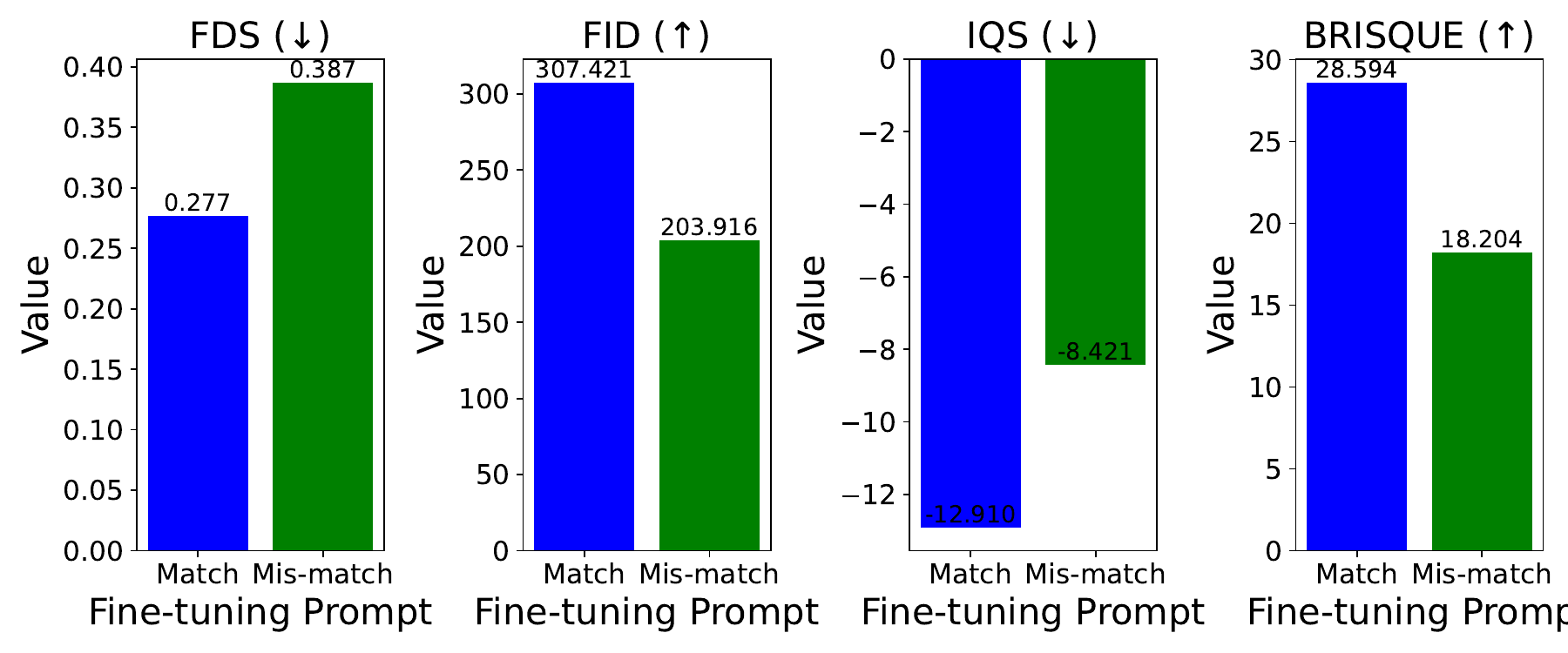}
    \caption{FSGM}
    \label{figure 2-1}
    \end{subfigure}
    \begin{subfigure}[t]{0.45\textwidth}
    \centering
        \includegraphics[width=1.0\textwidth]{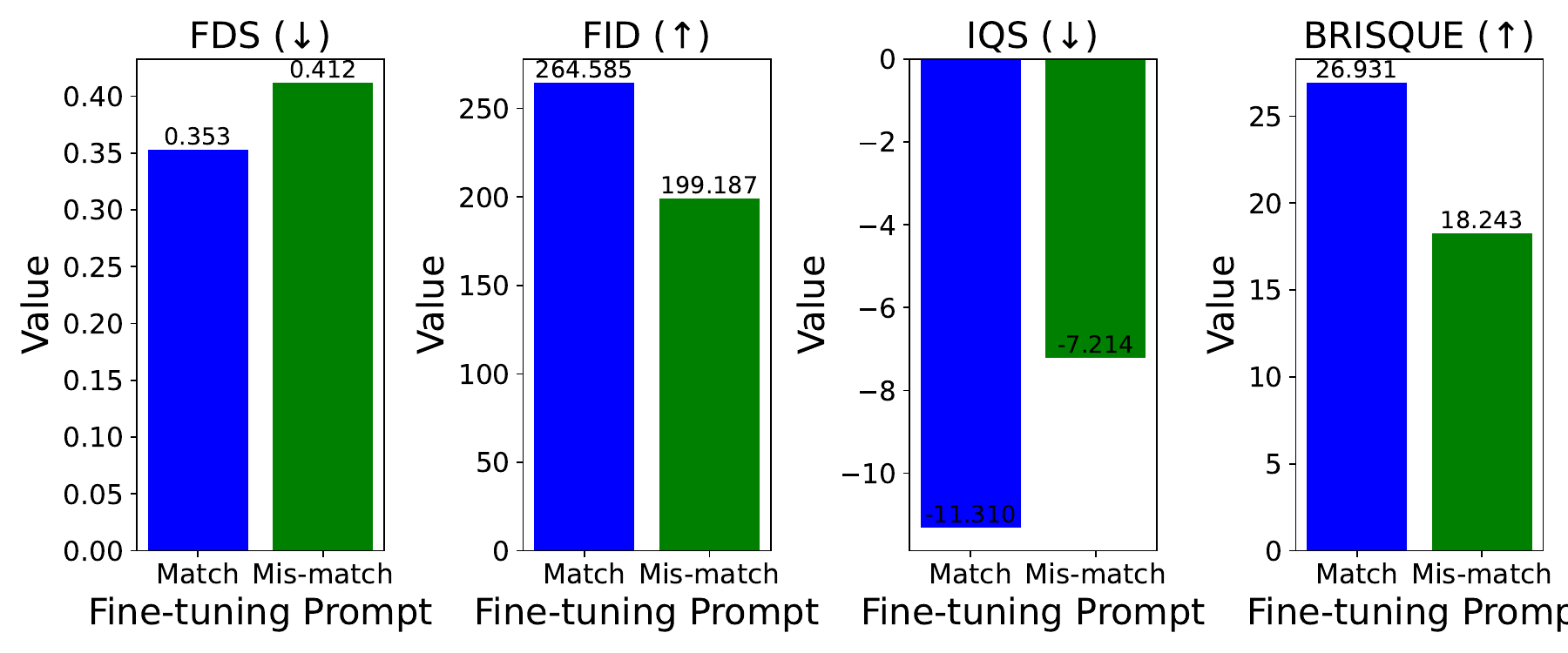}
    \caption{ASPL}
    \label{figure 2-2}
    \end{subfigure}
    \caption{The quantitative results showing the performance of the prompt-related defenses when textual prompts between the protection stage and the exploration stage are matched ($\bm{c_{prot}}=\bm{c_{explo}}$) and mismatched ($\bm{c_{prot}} \neq \bm{c_{explo}}$). }
    \label{figure2}
\end{figure}

\par \textbf{Results:} For the selected 4 images of each celebrity, we adopt the defense method FSGM and ASPL with the protecting prompt $\bm{c_{prot}}$ to generate the corresponding protected version. These protected images are then used to fine-tune the model with the fine-tuning prompt $\bm{c_{explo}}$, resulting in different fine-tuned models\footnote{When $\bm{c_{prot}} \neq \bm{c_{explo}}$, we choose the fine-tuned model that has the highest FDS.}. For testing, we use arbitrary prompts to generate a set of images, which are then evaluated using the four metrics mentioned above. The average results across different fine-tuned models are shown in Figure \ref{figure2}. We can see that the protection performance is notably affected when the protecting prompt differs from the fine-tuning prompt. For example on the FSGM method, when the fine-tuning prompts do not match the protecting prompts, the metric FDS increases over 35\% (0.277 $\to$ 0.387) and the metric FID decreases 30\% (307.421 $\to$ 203.916). The phenomenon is consistent for other metrics and methods.

Deeply concerned by our observation that breaking the prompt-consistency assumption made by the data protectors could enable the exploiters to generate high-quality mimic images, even when the data is safeguarded to some extent, we aim to design a prompt-agnostic defense in the following parts.

\section{Does Perturbing the Visual Encoer Affect Fine-tuning?}

\label{Sec:Latent-Space}

\par Recall that the latent distribution is modeled by a KL-based VAE \cite{vae} as a multinomial Gaussian Distribution, $\gN(\mu_\gE(\bm{x}), \sigma_\gE^2(\vx))$, which is prompt-independent. This property can be leveraged to address the defense degradation when there is a prompt mismatch. Before delving into this potential solution, we first investigate how changes in the latent distribution, i.e., in the mean $\mu_\gE(\bm{x})$ and the variance $\sigma_\gE^2(\bm{x})$, influence fine-tuning.
\par We define $L_{mean}$ to maximize the distance between the mean of the perturbed images and the clean images, while $L_{var}$ to maximize the distance between the variances of the two distributions. Formally, $L_{mean}$ and $L_{var}$ are  
\begin{align}
    L_{mean}(\bm{x}, \bm{\delta}) &= || \mu_\gE(\bm{x} + \bm{\delta}) - \mu_\gE(\bm{x})||_2^2, \\
    L_{var}(\bm{x}, \bm{\delta}) &= || \sigma_\gE(\bm{x} + \bm{\delta}) - \sigma_\gE(\bm{x})||_2^2,
\end{align}
where $\bm{\delta}$ denotes the perturbation added. We maximize the above loss functions with $\ell_{\infty}$-PGD$_{1000}$ \cite{pgd} and $\bm{\delta}$ is constrained by $\norm{\bm{\delta}}_\infty \leq \epsilon=0.05$. Then we conduct fine-tuning on the images obtained by optimizing the above two targets and evaluate the fine-tuned models with the same evaluation framework in Section \ref{Sec3}.

\begin{table}[!t]
\center
\caption{Evaluation of fine-tuning on the images which maximize the mean distance $L_{mean}$ and the variance distance $L_{var}$ respectively. \textit{Random} denotes adding random noise uniformly  within $[-\epsilon, \epsilon]$ to the clean image.}
\label{table1}
\begin{adjustbox}{width=.45\textwidth}    
\begin{tabular}{ c | c c c c } 
\toprule
Data & FDS $(\downarrow)$ & FID $(\uparrow)$ & IQS $(\downarrow)$ & BRISQUE $(\uparrow)$\\
\midrule
Clean & 0.480 & 144.570 & 4.310 & 15.447 \\
Random & 0.479 & 150.788 & 4.504 & 12.160 \\
$L_{mean}$ & 0.370 & 243.292 & \textbf{-5.373} & \textbf{21.655} \\
$L_{var}$ & \textbf{0.329} & \textbf{265.337} & -0.926 & 16.369 \\
\bottomrule 
\end{tabular}%
\end{adjustbox}%
\end{table} %

\begin{figure}[!t]
\centering
\vspace{\floatsep} %
\begin{subfigure}[b]{0.10\textwidth}
    \centering
    \includegraphics[width=\linewidth]{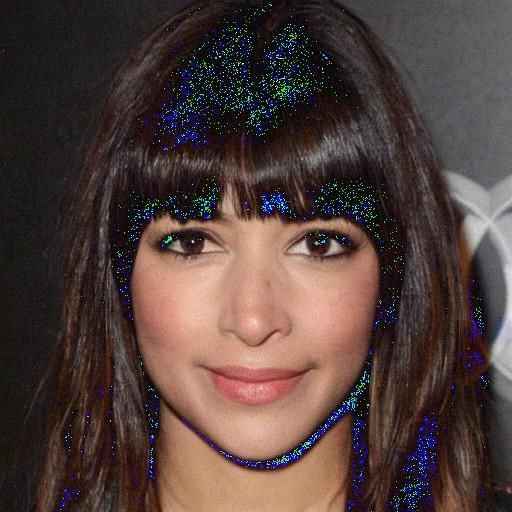}
    \caption{0 steps}
    \label{fig3-1}
\end{subfigure}
~
\begin{subfigure}[b]{0.10\textwidth}
    \centering
    \includegraphics[width=\linewidth]{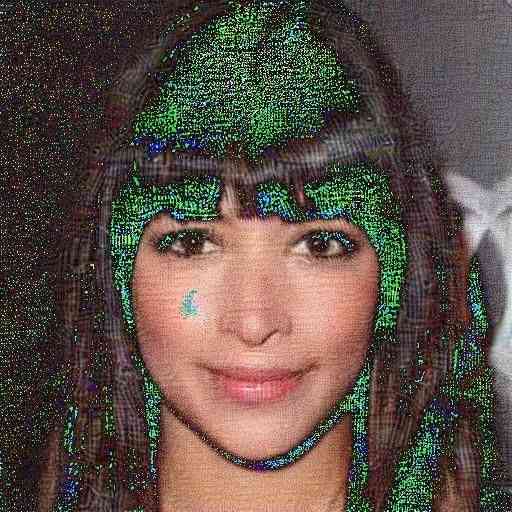}
    \caption{300 steps}
    \label{fig3-2}
\end{subfigure}
~
\begin{subfigure}[b]{0.10\textwidth}
    \centering
    \includegraphics[width=\linewidth]{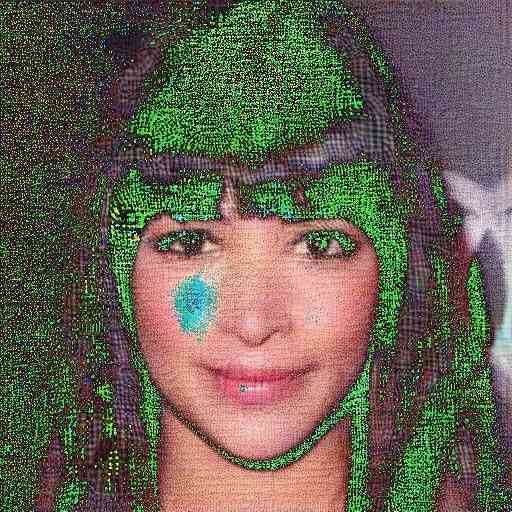}
    \caption{600 steps}
    \label{fig3-3}
\end{subfigure}
~
\begin{subfigure}[b]{0.10\textwidth}
    \centering
    \includegraphics[width=\linewidth]{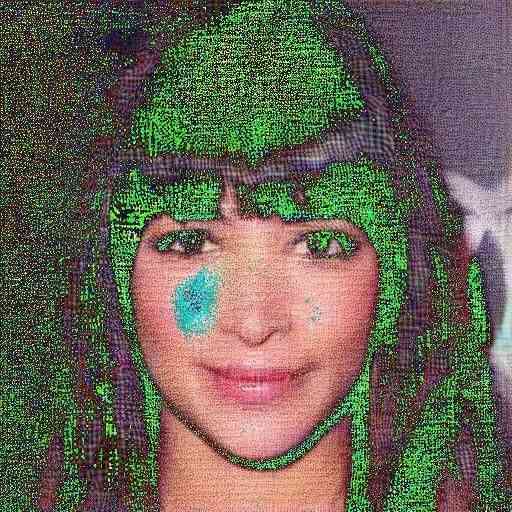}
    \caption{900 steps}
    \label{fig3-4}
\end{subfigure}

\vspace{\floatsep} %
\begin{subfigure}[b]{0.10\textwidth}
    \centering
    \includegraphics[width=\linewidth]{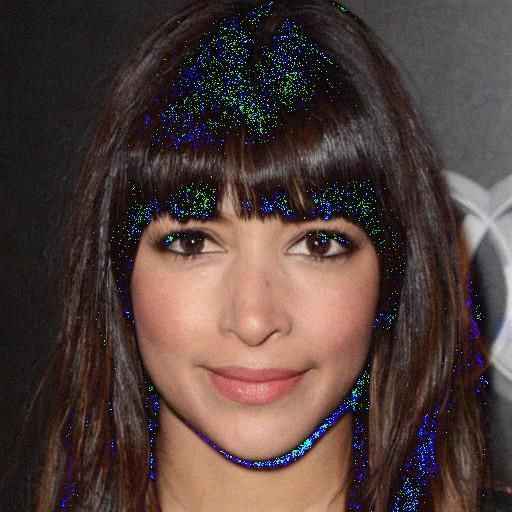}
    \caption{0 steps}
    \label{fig3-5}
\end{subfigure}
~
\begin{subfigure}[b]{0.10\textwidth}
    \centering
    \includegraphics[width=\linewidth]{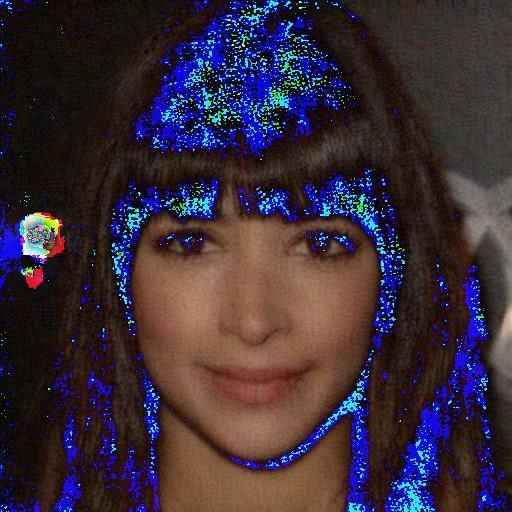}
    \caption{300 steps}
    \label{fig3-6}
\end{subfigure}
~
\begin{subfigure}[b]{0.10\textwidth}
    \centering
    \includegraphics[width=\linewidth]{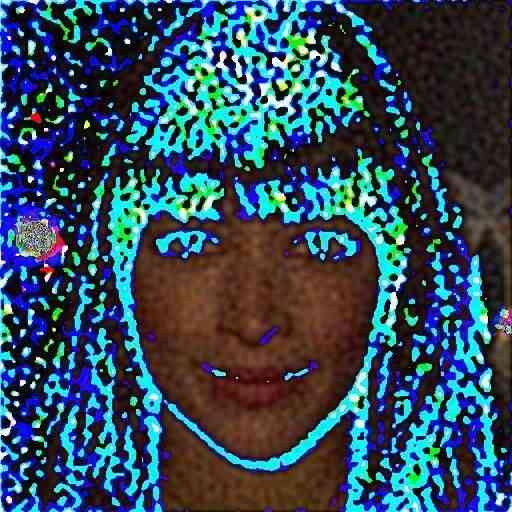}
    \caption{600 steps}
    \label{fig3-7}
\end{subfigure}
~
\begin{subfigure}[b]{0.10\textwidth}
    \centering
    \includegraphics[width=\linewidth]{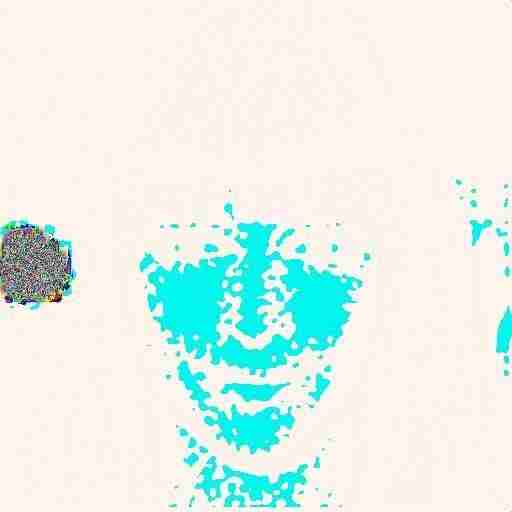}
    \caption{900 steps}
    \label{fig3-8}
\end{subfigure}
\caption{Visualizations of the perturbed latent representations. We decode the latent representations $z$ obtained during the maximization of $L_{mean}$ and $L_{var}$ with the visual decoder in the LDM. (a) to (d) corresponding to the change of mean, while (e) to (h) for the variance. The images are obtained with the Stable Diffusion v1.5.}
\label{figure3}
\end{figure}

\par The results presented in Table \ref{table1} demonstrate that significantly reshaping the latent distribution indeed has a substantial influence on fine-tuning. To visually illustrate the influence of the distorted latent distribution, we decode the representations $\bm{z}$ sampled from the distribution during the optimization process using the visual decoder and display the decoded images in Figure \ref{figure3}\footnote{Note that, for this specific experiment, we directly input ($\bm{x} + \bm{\delta}$) as float numbers to the encoder without converting to uint-8. This approach maximally showcases the perturbations' influence.}. Combing results in Table \ref{table1} and Figure \ref{figure3}, we find that a large mean difference with the clean images mainly influences the texture of the output images, making them appear covered with heavy noise (low IQS and high BRISQUE). Conversely, a large variance significantly prohibits the model from grasping the core concepts of the images (low FDS and high FID).

\par
\begin{figure}[!t]
    \vspace{-1mm}
    \centering 
    \begin{subfigure}[b]{0.24\textwidth}
        \centering
        \includegraphics[width=1.0\textwidth]{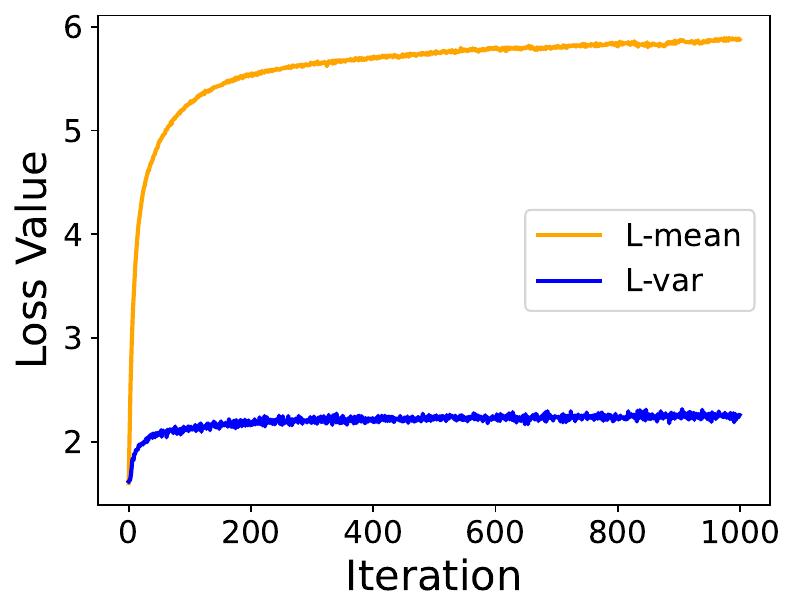}
        \caption{$||\mu_\gE(x + \delta)-\mu_\gE(x)||_2^2$}
        \label{figure4-1}
    \end{subfigure}%
    ~
    \begin{subfigure}[b]{0.24\textwidth}
        \centering
        \includegraphics[width=1.0\textwidth]{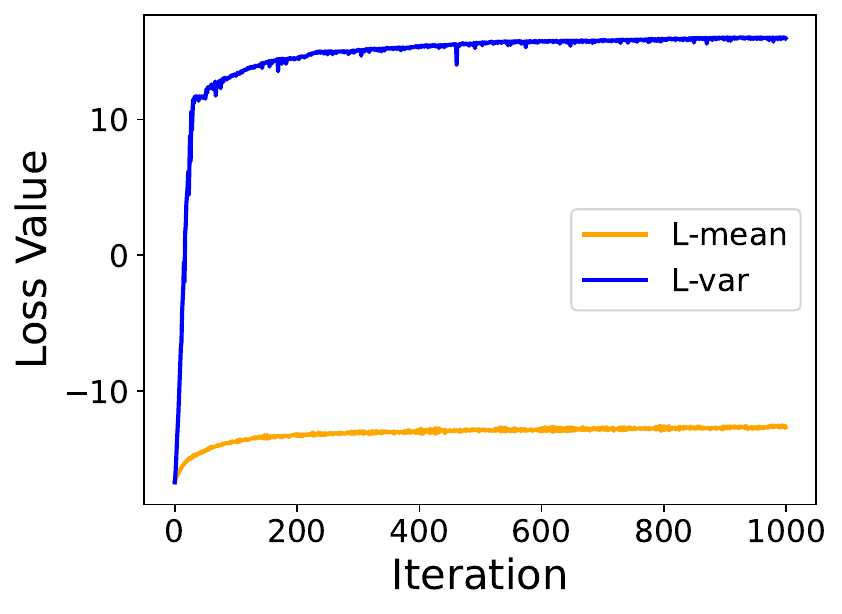}
        \caption{$\log \sigma_\gE(x + \delta)^2 - \log \sigma_\gE(x)^2$}
        \label{figure4-2}
    \end{subfigure}
    
    \caption{The change of the latent distribution as the perturbations are added. (a) the change of the $\ell_2$ distance between the mean of the clean and the perturbed latent distribution. (b) the change of the $\ell_2$ distance between the variance of the clean and the perturbed latent distribution. The target of each colored line is shown in the figure legend.}
    \vspace{-1mm}
    \label{figure4}
\end{figure}

 Lastly, we plot the $\ell_2$ norm of the mean difference and the variance difference in Figure \ref{figure4}, which reveals that even small perturbations added in the pixel space (0.05) can significantly alter the latent distribution. The variance changes so drastically that the gap between the variance of clean images and that of perturbed images ranges from approximately $\sim 2^{-15}$ to $\sim 2^{12}$. Additionally, we observe that changes in the mean and changes in the variance are not entirely correlated. Whatever in Figures \ref{figure4-1} or \ref{figure4-2}, one undergoes significant fluctuations while the other does not exhibit substantial variation, which indicates their distinct impacts on fine-tuning results. 

Overall, by inducing perturbations in the pixel space, we can manipulate the two statistics of the latent distribution, thereby significantly affecting different aspects of the fine-tuning outcome.

\section{How Can We Make Better Use of the Visual Encoder for Data Protection?}
As discussed in Section \ref{Sec:Latent-Space}, perturbing the latent distribution significantly impacts the fine-tuning process, and notably, this latent distribution is prompt-independent. Therefore, in this section, we aim to utilize the visual encoder to implement an effective prompt-independent defense mechanism.

\subsection{Proposed Prompt-Independent Defense (PID)}
\par From the results in Table \ref{table1}, we know that influencing the mean and variance impacts different aspects of the learning procedure. Observing Figure \ref{figure4}, we know that altering just one of these statistics is insufficient to simultaneously induce substantial changes in both.
This observation triggers us to explore the possibility of manipulating the latent distribution more effectively by designing a sophisticated target that leverages the benefits of influencing both the mean and the variance. 

We first attempt to disrupt the representations $\bm{z} = \gE(\bm{x}, \bm{\varepsilon})$ sampled from the latent distribution, leading to the loss function $L_{sample}$, which is also employed by \citet{advDM}. To reduce unnecessary randomness in the optimization process, we then experiment with excluding $\bm{\varepsilon}$ from $L_{sample}$, resulting in the loss function $L_{add}$. Considering the significant disparity in the magnitudes of the mean and variance ($10^2$ vs $10^{-3}$) observed in Figure \ref{figure4}, we propose $L_{add-log}$, which jointly optimizes the logarithm of the variance and the mean. Additionally, we explore targeted manipulation of the mean $\bm{x_{target}}$, as done by \citet{liang2023mist}, where the target is the default image specified in their paper\footnote{https://github.com/mist-project/mist/blob/main/resources}. We denote this loss as $L_{mean}^T$.
\begin{align}
    L_{sample}(\bm{x}, \bm{\delta}) &= \E_{\bm{\varepsilon_1}, \bm{\varepsilon_2}} ||\gE(\bm{x + \delta}, \bm{\varepsilon_1}) - \gE(\bm{x}, \bm{\varepsilon_2})||_2^2, \\  
    L_{add}(\bm{x}, \bm{\delta}) &= L_{mean} + L_{var}, \\
    L_{add-log}(\bm{x}, \bm{\delta}) &= L_{mean} + \log \frac{\sigma_\gE(\bm{x} + \bm{\delta})^2}{\sigma_\gE(\bm{x})^2}, \\
    L^T_{mean}(\bm{x}, \bm{\delta}) &= -|| \mu_\gE(\bm{x} + \bm{\delta}) - \mu_\gE(\bm{x_{target}})||_2^2,
\end{align}

\par
\begin{figure}[!t]
    \centering 
    \begin{subfigure}[b]{0.24\textwidth}
        \centering
        \includegraphics[width=1.0\textwidth]{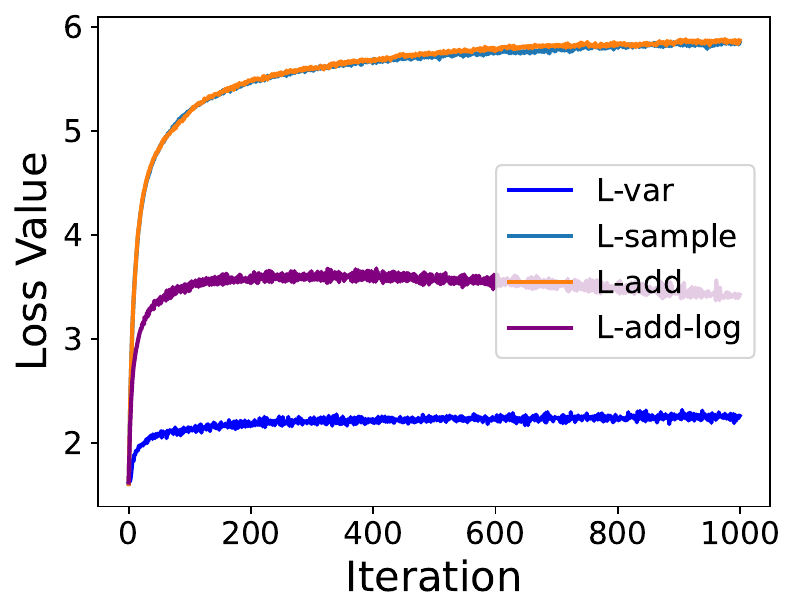}
        \caption{$||\mu_\gE(x + \delta)-\mu_\gE(x)||_2^2$}
        \label{fig5-1}
    \end{subfigure}%
    ~
    \begin{subfigure}[b]{0.24\textwidth}
        \centering
        \includegraphics[width=1.0\textwidth]{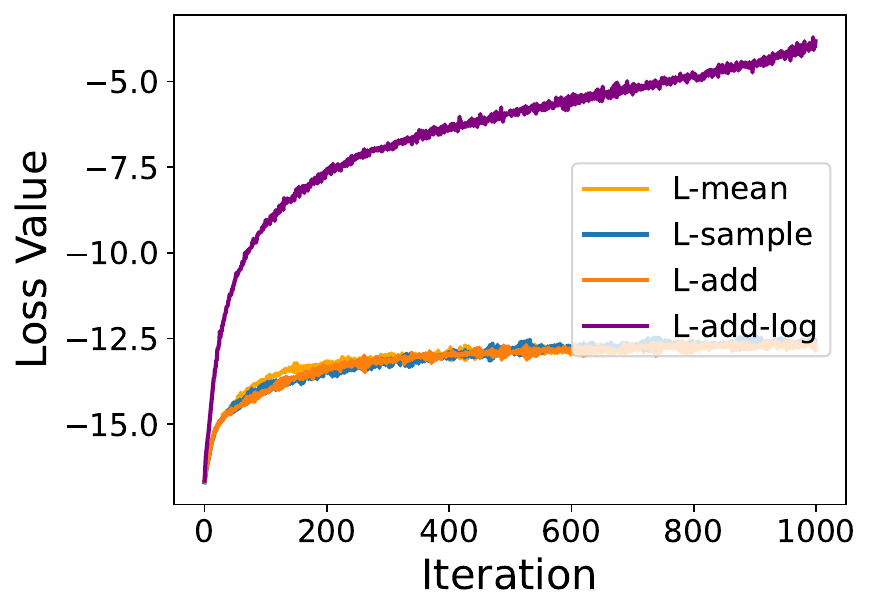}
        \caption{$\log \sigma_\gE(x + \delta)^2 - \log \sigma_\gE(x)^2$}
        \label{fig5-2}
    \end{subfigure}
    \caption{The change of the latent distribution as the perturbations are added. $L_{add-log}$ is the only loss that has a significant impact on both statistics. (a) and (b) We plot the change of the mean and the variance of perturbed images as the maximization of each loss goes respectively. The results are averaged over all elements in the tensor.}
    \vspace{-1mm}
    \label{fig5}
\end{figure}

 We proceed by evaluating the influence of the defense targets proposed above on the latent distributions, as done in Figure \ref{figure4}. Notably, we notice that $L_{add-log}$ (the purple line in Figure \ref{fig5-1} and Figure \ref{fig5-2}) is the only defense target that shifts both statistics away from their normal values significantly with averaged $\ell_2$ distance of the mean being 3.5 and 0.06 for variance.
On the contrary, $L_{sample}$ and $L_{add}$ perform significantly worse in perturbing variance. 

Equipped with a suitable target, we next examine whether it has a larger influence on fine-tuning than before. The results presented in Table \ref{table2} reveal that the potential of the encoders in data protection has not been fully explored before. The loss functions adopted by previous literature, $L_{mean}$, $L_{mean}^T$, and $L_{add}$, exhibit sub-optimal performance compared to $L_{add-log}$. We note that the similar behaviors of $L_{mean}$, $L_{sample}$, and $L_{add}$ can be well explained by observations in Figure \ref{fig5-1} and Figure \ref{fig5-2}, as all of them mostly focus on the mean value.

\par A carefully designed optimization target, $L_{add-log}$, proves to combine the advantages of influencing $\mu$ and $\sigma$. Not only does it successfully stop the model from learning the human face (low FDS, high FID), but it also notably affects the structure and texture of the output images (low IQS and high BRISQUE). We are surprised to find that $L_{add-log}$ even outperforms both $FSGM (c=c_1)$ and $ASPL (c=c_1)$ under this fine-tuning configuration. Given that the latter defenses require much more GPU memory since they involve the UNet \cite{unet} model, which is much heavier than the visual encoder, we thus believe the visual encoder should play an undiminished role in protecting data against LDMs. For its superior protection effect and independence of textual conditions, we implement the defense target defined by $L_{add-log}$ as the \textbf{Prompt-Independent Defense (PID)}.

\begin{table}[!t]
\centering
\vspace{-1mm}
\caption{Evaluation of fine-tuning on the images maximizing the losses defined from Equation (8) to (11), together with $L_{mean}^T$. The fine-tuned model is Stable Diffusion v1.5.}
\label{table2}
\begin{adjustbox}{width=.45\textwidth}    
\begin{tabular}{ c | c c c c } 
\toprule
Data & FDS $(\downarrow)$ & FID $(\uparrow)$ & IQS $(\downarrow)$ & BRISQUE $(\uparrow)$\\
\midrule
Clean & 0.480 & 144.570 & 4.310  & 15.447 \\
Random & 0.479 & 150.788 & 4.504  & 12.160 \\
$L_{mean}^T$ & 0.377 & 271.540 & -4.047 & 28.622 \\
$L_{sample}$ & 0.377 & 265.588 & -5.135 & 21.119 \\
$L_{add}$ & 0.377 & 268.260 & -5.500 & 20.465 \\
$L_{add-log}$ & \textbf{0.329} & \textbf{411.990}  & \textbf{-18.296} & \textbf{35.510} \\
\bottomrule
\end{tabular}%
\end{adjustbox}%
\vspace{-1mm}
\end{table} %

\subsection{Integrating PID with Existing Defenses}
\par We continue to explore the possibility of improving the current defenses with PID. To combine the two distinct types of defense, namely defense with the encoder and defense with attacking the training loss function, we adopt a joint optimization approach involving a weighted combination of both two defense objectives, similar to \citet{advDM} and \citet{liang2023mist}. Specifically, given a defense target $T$ that incorporates the training loss of LDMs and a defense target aimed at manipulating the latent distribution, $L$, we define a tradeoff coefficient $\lambda$ to balance the two targets. The combined defense is expressed as follows:
\begin{equation}
    L_{combo}(\theta, c, \bm{x}) = T(\theta, c, \bm{x}) + \lambda L(\bm{x}).
\end{equation}
Here, $\theta$ again denotes the model parameter, $c$ represents the textual condition, and $\bm{x}$ is the data to protect. $T(\theta, c, \bm{x})$ can be the defense methods defined in equation \ref{FSGM-Defense} and equation \ref{ASPL-Defense}.
\par We let $L=L_{add-log}$ the strongest defense target we observed, and $T \in$ \{ASPL, FSGM\}. We empirically identify $\lambda=0.05$ as the best parameter in our default setting. While it's possible to exhaustively search for the optimal $\lambda$ across all settings, we consistently adopt $\lambda^*=0.05$ in the following experiments due to the massive computational demands.

\begin{table*}[h]
\vspace{-1mm}
\centering
\caption{Evaluation of defense algorithms under a controlled scenario $\bm{c_{prot}} = \bm{c_{explo}}$ and a more realistic scenario where $\bm{c_{prot}} \neq \bm{c_{explo}}$ on the CelebA-HQ dataset. The best-performing defense under each metric is marked with \textbf{bold}. Frozen text-encoder means we freeze the parameters of the text-encoder during fine-tuning, while unfrozen means we simultaneously fine-tune the text-encoder and UNet. }
\label{table3}
\begin{subtable}[t]{0.95\textwidth}
\centering
\begin{adjustbox}{width=1.0\textwidth}
\begin{tabular}{ c | c | c  c c c c | c c c c }
\toprule
\multirow{2}{*}{Scenario}& \multirow{2}{*}{Data} & \multicolumn{4}{c}{Frozen Text Encoder} &&   \multicolumn{4}{c}{Unfrozen Text Encoder} \\
\cline{3-4} \cline{5-11}
& &  FDS$(\downarrow)$ & FID$(\uparrow)$ & IQS$(\downarrow)$ & BRISQUE$(\uparrow)$ && FDS$(\downarrow)$ & FID$(\uparrow)$ & IQS$(\downarrow)$ & BRISQUE$(\uparrow)$ \\
\midrule
     & Clean & 0.482 & 144.570 & 4.397 & 14.757 && 0.557 & 152.870 & 7.104 & 18.445 \\
     \midrule
        \multirow{4}{*}{$\bm{c_{prot}} = \bm{c_{explo}}$} 
        & AdvDM & 0.344 & 240.452 & -11.310 & 19.100 && 0.358 & 208.859 & -8.558 & 25.472 \\
        & FSGM & 0.342 & 246.434 & -8.710 & 22.046 && 0.277 & 307.421 & -12.910 & 28.594\\
        & ASPL & 0.330 & 295.415 & -9.558 & 26.993 && 0.353 & 264.585 & -11.310 & 26.931\\
        & PID  & \textbf{0.205} & \textbf{411.990} & \textbf{-18.296} & \textbf{49.178} && \textbf{0.257} & \textbf{325.962} & \textbf{-29.693} & \textbf{62.749} \\
        
    \cmidrule{1-11}
    \multirow{4}{*}{$\bm{c_{prot}} \neq \bm{c_{explo}}$} 
        & AdvDM & 0.378 & 253.501 & -8.534 & 16.692 && 0.407 & 216.154 & \textbf{-9.260} & 21.813\\
        & FSGM & 0.374 & 224.460 & -5.607 & 20.678 && 0.387 & 203.916 & -8.421 & 18.204 \\
        & ASPL & 0.370 & 185.074 & -3.669 & 26.993 && 0.412 & 199.187 & -7.214 & 18.243\\
        & PID  & \textbf{0.254} & \textbf{352.795} & \textbf{-15.273} & \textbf{35.510} &&  \textbf{0.303} & \textbf{307.760} & -8.979 & \textbf{28.927}\\
\bottomrule
\end{tabular}
\end{adjustbox}
\subcaption{The fine-tuned model is Stable Diffusion v1.5.}
\label{table3-1}
\vspace{-1mm}
\end{subtable}%

\begin{subtable}[t]{0.95\textwidth}
\centering
\begin{adjustbox}{width=1.0\textwidth}
\begin{tabular}{ c | c | c  c c c c | c c c c }
\toprule
\multirow{2}{*}{Scenario}& \multirow{2}{*}{Data} & \multicolumn{4}{c}{Frozen Text Encoder} &&   \multicolumn{4}{c}{Unfrozen Text Encoder} \\
\cline{3-4} \cline{5-11}
& &  FDS$(\downarrow)$ & FID$(\uparrow)$ & IQS$(\downarrow)$ & BRISQUE$(\uparrow)$ && FDS$(\downarrow)$ & FID$(\uparrow)$ & IQS$(\downarrow)$ & BRISQUE$(\uparrow)$ \\
\midrule
     & Clean & 0.494 & 175.856 & 9.959 & 11.654 && 0.565 & 140.735 & 5.641 & 10.199 \\
     \midrule
        \multirow{4}{*}{$\bm{c_{prot}} = \bm{c_{explo}}$} 
        & AdvDM & 0.322 & 252.407 & -2.127 & 24.382 && 0.362 & 229.829 & -12.287 & 31.833\\
        & FSGM & 0.298 & 277.588 & 0.586 & 30.837 && 0.312 & 257.165 & -7.456 & 30.764 \\
        & ASPL & 0.300 & 282.938 & 0.012 & 31.429 && 0.313 & \textbf{266.097} & -5.707 & 28.832 \\
        & PID  & \textbf{0.255} & \textbf{350.382} & \textbf{-16.556} & \textbf{50.757} && \textbf{0.288} & 260.496 & \textbf{-14.764} & \textbf{50.112} \\
        
    \cmidrule{1-11}
    \multirow{4}{*}{$\bm{c_{prot}} \neq \bm{c_{explo}}$} 
        & AdvDM & 0.346 & 245.780 & -5.081 & 24.293 && 0.398 & 231.861 & -10.128 & 31.006 \\
        & FSGM & 0.326 & 252.407 & 2.365 & 30.950 && 0.347 & 234.313 & -6.279 & 24.918 \\
        & ASPL & 0.341 & 236.257 & -1.872 & 30.717 && 0.388 & 203.413 & -1.541 & 23.357 \\
        & PID  & \textbf{0.285} & \textbf{336.617} & \textbf{-12.634} & \textbf{43.746} && \textbf{0.288} & \textbf{366.596} & \textbf{-14.764} & \textbf{50.112}\\
\bottomrule
\end{tabular}
\end{adjustbox}
\subcaption{The fine-tuned model is Stable Diffusion v2.1.}
\label{table3-2}
\end{subtable}%
\end{table*}

\begin{table*}[h]
\centering
\caption{Evaluation of PID-hybridized defense algorithms under an uncontrolled scenario where $\bm{c_{prot}}\neq \bm{c_{explo}}$. The default training prompt \textit{A photo of sks person} is adopted as $\bm{c_{prot}}$. The backbone model is Stable Diffusion v1.5 and $\lambda^*=0.05$. The superior one between FSGM/ASPL and FSGM+PID/ASPL + PID under each metric is marked with \textbf{bold}.}
\label{table4}
\begin{subtable}[t]{0.95\textwidth}
\begin{adjustbox}{width=1.0\textwidth}    
\begin{tabular}{ c | c c c c c | c c c c }
\toprule
\multirow{2}{*}{Data}& \multicolumn{4}{c}{Frozen Text Encoder} & &  \multicolumn{4}{c}{Unfrozen Text Encoder} \\
\cline{2-5} \cline{6-10}
& FDS $(\downarrow)$ & FID $(\uparrow)$ & IQS $(\downarrow)$ & BRISQUE $(\uparrow)$ & & FDS $(\downarrow)$ & FID $(\uparrow)$ & IQS $(\downarrow)$ & BRISQUE $(\uparrow)$ \\
\midrule
FSGM & 0.374 & \textbf{224.460} & \textbf{-5.607} & 20.678 && 0.387 & \textbf{203.916} & \textbf{-8.421} & 18.204 \\
FSGM + $\lambda^*$PID & \textbf{0.276} & 185.704 & -3.669 & \textbf{21.096} && \textbf{0.303} & 185.074 & -3.665 & \textbf{19.096}\\ 
\midrule
ASPL & 0.370  & 271.893 & -5.786 & 22.724 && 0.412 & 199.187 & \textbf{-7.214} & 18.243\\
ASPL + $\lambda^*$PID& \textbf{0.254} & \textbf{352.795} & \textbf{-15.723} & \textbf{35.510} && \textbf{0.335} & \textbf{208.859} & -3.443 & \textbf{20.659} \\
\bottomrule
\end{tabular}
\end{adjustbox}
\end{subtable}%
\vspace{-1mm}
\end{table*} %

\section{Experiments}
\label{Sec6}
In this part, we first assess the performance of PID as well as existing algorithms under both the $\bm{c_{prot}}=\bm{c_{explo}}$ and the $\bm{c_{prot}} \neq \bm{c_{explo}}$ scenarios. Second, we experiment with combining PID and existing defenses. Last but not least, we test the robustness of PID under harsh conditions.

\subsection{PID Excels Regardless of the Prompt Consistency Assumption}
\par \textbf{Experiment Setup: }We compare PID with the three symbolic defense methods, AdvDM \cite{advDM}, FSGM, and ASPL \cite{antidreambooth} on the CelebA-HQ \cite{celeba} and VGGFACE \cite{cao2018vggface2} dataset. We largely adopt their default configurations when running the defense algorithms. We generate PID with PGD$_{1000}$ \cite{pgd} and the perturbation budget is set to $\epsilon_\infty = 0.05$. We use the Stable Diffusion v1.5 and Stable Diffusion v2.1 \cite{stablediffusion} as the base model\footnote{We obtain the Stable Diffusion v1.5 from \hyperlink{v1.5}{https://huggingface.co/runwayml/stable-diffusion-v1-5} and the Stable Diffusion v2.1 from \hyperlink{v2.1}{https://huggingface.co/stabilityai/stable-diffusion-2-1.}}. The evaluation protocol is identical to the one introduced in Section \ref{Sec3} with the experimental details provided in Appendix \ref{Appendix A}. 

\begin{figure}[h]
\centering
\vspace{-1mm}
    \includegraphics[width=0.45\textwidth]{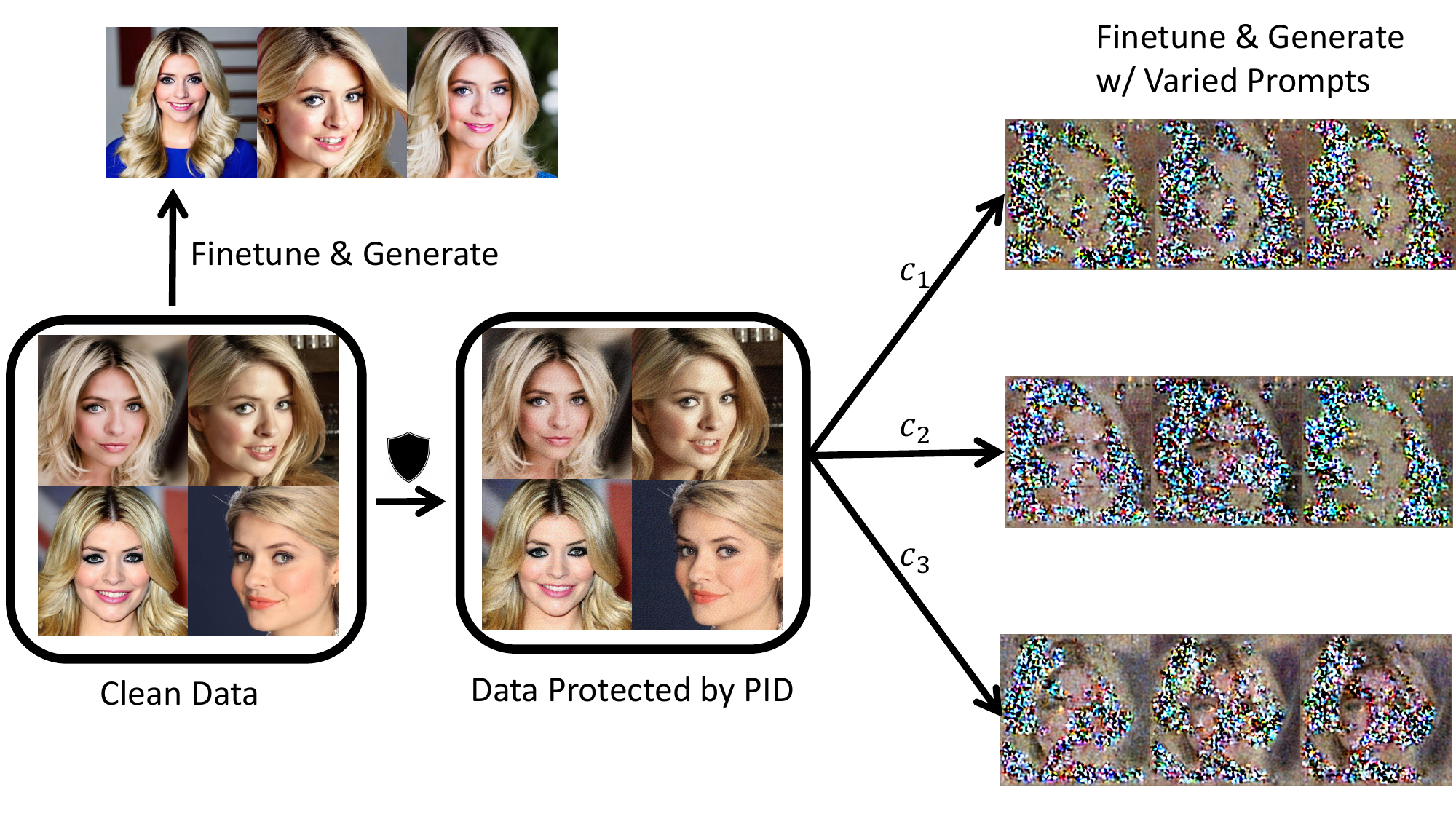}
    \caption{Illustration of PID's defense performance against varied prompts used by the data exploitors. $\bm{c_1}$ to $\bm{c_3}$ correspond to three different fine-tuning prompts. In the figure, we finetune the SD v1.5. More visualizations can be found in Appendix \ref{AppendixE}.}
    \label{fig6}
\vspace{-1mm}
\end{figure}
\par \textbf{Results: }The complete results for CelebA-HQ are listed in Table \ref{table3}. Remarkably, despite consuming significantly less computational resources (approximately 20\% GPU memory, 5G v.s. 24G), PID achieves comparable, if not superior, performance compared to the three algorithms incorporating UNet across all four training configurations.
Specifically, when the text encoder is frozen, i.e., not trained, during fine-tuning, PID consistently prohibits the LDMs from learning useful semantical information, resulting in notably poor facial similarity (0.254 for SD v1.5 and 0.285 for SD v2.1). Regarding the case where the text encoders are also fine-tuned, PID induces severely noisy, low-quality images that have little semantical correlation with the training data, as evidenced by the degraded FDS (0.303 and 0.288), substantially reduced IQS (-8.979 and -14.764), and high BRISQUE (28.927 and 50.112). Benefiting from the visual encoders' independence from the text encoders, PID consistently results in FID greater than 300 across all settings, rendering the generated images unrelated to the training data.
The results demonstrate the potential for PID as a strong baseline method for safeguarding data against LDMs regardless of the varied prompts. The results for VGGFACE \cite{cao2018vggface2} and LoRA finetuning \cite{lora} are provided in Appendix \ref{AppendixC} and we showcase images generated by the LDMs fine-tuned on the protected data in Appendix \ref{AppendixE}.

\subsection{Hybriding PID Robustifies Current Algorithms}
We then compare the prompt-dependent defenses with their PID-hybridized variants. Results in Table \ref{table4} reveal that PID is able to enhance the robustness of the current algorithms. 
we observe that ASPL+PID is much more robust than ASPL regardless of whether the text encoder is frozen or not, as supported by the notably lower FDS (0.254 v.s. 0.370, 0.335 v.s. 0.412) and higher FID (352 v.s. 271, 208 v.s. 199). Moreover, FSGM+PID consistently results in images of worse semantic information than FSGM (lower FDS). Based on the above results, we argue that PID can be incorporated into existing defenses for more reliable data protection against LDMs.
\par We do not ignore the fact that combining PID with FSGM fails to do better in image quality, which might be attributed to a sub-optimal $\lambda^*$ choice or the difficulty of joint optimization. Though inferior, FSGM+PID still qualifies for a valid defense for its larger influence on semantic information (lower FDS). 

\subsection{Improved Cross-model Transferability}
\par Since the data protectors have no control over what model the downstream exploiter uses, it is possible that different models are adopted in the two stages. We examine the transferability between different models of PID as well as existing algorithms under the $\bm{c_{prot}}\neq \bm{c_{explo}}$ setting. Specifically, we consider the transferability between the Stable Diffusion v1.5 and Stable Diffusion v2.1.
\par \textbf{Results: } PID enjoys great transferability between the two model versions as shown in Table \ref{table5}, which might be due to the similarity in the condensed representations of images. Also, we notice that the transferability of existing algorithms from SD v2.1 to SD v1.5 is relatively weaker than the other way around.

\begin{table}[h]
\vspace{-1mm}
\centering
\caption{The transferability of different data protection algorithms. The images are protected with the \textit{Source} models and are exploited by the \textit{Target} models. v1.5 and v2.1 denote Stable Diffusion v1.5 and Stable Diffusion v2.1 repectively. The text encoders are frozen when fine-tuning the models in this table.}
\label{table5}
\begin{adjustbox}{width=.45\textwidth}
\begin{tabular}{c | c | c c c c }
\toprule
Src.$\to$Dst. & Data & FDS $(\downarrow)$ & FID $(\uparrow)$ & IQS $(\downarrow)$ & BRISQUE $(\uparrow)$ \\
\midrule
\multirow{5}{*}{v2.1 $\to$ v1.5} 
& AdvDM & 0.371 & 223.914 & -2.083 & 30.010\\
& FSGM &0.364 & 208.278 & -2.339 & 37.563 \\
& ASPL & 0.311 & 252.740 & -2.510 & 37.706 \\
& PID & \textbf{0.268} & \textbf{350.069} & \textbf{-14.802} & \textbf{46.204}\\
\midrule
\multirow{5}{*}{v1.5 $\to$ v2.1} 
& AdvDM & 0.407 & 231.139 & -4.660 & 17.108 \\
& FSGM & 0.372 & 241.951 & -7.724 & 23.091 \\
& ASPL & 0.397 & 239.222 & -6.606 & 23.950 \\
& PID & \textbf{0.265} & \textbf{251.253} & \textbf{-15.087} & \textbf{24.365}\\
\bottomrule
\end{tabular}%
\end{adjustbox}%
\vspace{-1mm}
\end{table}

\subsection{Resilliance to Adaptive Attacks}
\par We continue studying the robustness of PID when faced with adaptive attacks with our quantitative results reported in Table \ref{table6}.
\par \textbf{Adaptive Attack: }Since our proposed PID focuses on manipulating the mean and the variance of the latent distribution, there could be adaptive attacks trying to break the conditions for our defense to be effective. We propose three possible adaptive attacks and test the robustness of our proposed defenses against them. 
\par \textbf{(1) Zero $\sigma$: }As it is shown in Figure \ref{fig5-2}, PID causes the variance of the latent distribution to increase dramatically. Therefore, the attack might fix the standard value of the latent distribution $\sigma_\gE(\vx)$ of the perturbed images to be $0$ to mitigate such effect. However, a zero standard value will make the finetuning process easier to overfit and lead to inferior generation results. Our results also reveal that PID works very well in such training settings with FDS$=0.253$ and IQS$=-9.313$.
\par \textbf{(2) Clipped $\sigma$ \& (3) Fixed $\sigma$: } A smarter attacker might try clipping or fixing the standard value $\sigma_\gE(x)$ to a relatively normal value, e.g. $10^{-7}$, rather than directly fixing it to be 0. Adopting the attack, we observe PID's influence on image quality is weakened, with the improved IQS and decreased FID shown in Table \ref{table6}. However, the FDS is still very low (< 0.3), rendering the attack ineffective.  
\par In all, PID is believed to be tolerant of the adaptive attacks we proposed above and exhibit convincing robustness.

\begin{table}[h]
\center
\vspace{-3mm}
\caption{Robustness of PID against three adaptive attacks we proposed. All evaluations are done via fine-tuning a Stable Diffusion v1.5. The best-performing attack is marked as \textbf{bold}.}
\label{table6}
\begin{adjustbox}{width=.45\textwidth}    
\begin{tabular}{c | c | c c c c }
\toprule
Freeze-TE & Data & FDS $(\downarrow)$ & FID $(\uparrow)$ & IQS $(\downarrow)$ & BRISQUE $(\uparrow)$ \\
\midrule
\multirow{4}{*}{\cmark} & Clean & 0.480 & 144.570 & 4.130 & 14.757 \\
& Zero $\sigma$ & \textbf{0.253} & \textbf{201.951} & -9.313 & 22.686 \\
& Clipped $\sigma$ & 0.249 & 207.611 & -12.968 & 33.156  \\
& Fixed $\sigma$  & 0.239 & 207.611 & \textbf{-6.238} & \textbf{22.685} \\
\midrule
\multirow{4}{*}{\xmark} & Clean & 0.557 & 128.870 & 7.104 & 18.445 \\
& Zero $\sigma$ & 0.257 & 228.788 & \textbf{-7.793} & 36.637 \\
& Clipped $\sigma$ & \textbf{0.279} & 367.174 & -16.602 & 37.565 \\
& Fixed $\sigma$ & 0.249 & \textbf{207.260} & -13.215 & \textbf{30.168} \\
\bottomrule
\end{tabular}%
\end{adjustbox}%
\vspace{-3mm}
\end{table} %

\subsection{Robustness against Data Corruptions}
\par After releasing the protected data, the protectors have no control over what data exploiters will do to the images. Here we consider four common data corruptions that may influence the effect of the protective perturbations, namely \textbf{randomly resizing and cropping}, \textbf{smoothing with uniform noise}, \textbf{image denoising} \footnote{We adopt the Gaussian image denoiser from the Aydin library, \hyperlink{aydin}{https://github.com/royerlab/aydin}} and \textbf{JPEG compression}. The model we used for this part is SD v1.5 and we freeze the text-encoder during fine-tuning. The reported experiments are done with $\bm{c_{prot}} = \bm{c_{explo}}$.
\par \textbf{Result: }Based on Table \ref{table7} we can observe that PID, the simplest defense among the four algorithms, withstands all four corruptions as evidenced by consistently low FDS and high FID. PID shows comparable performance to the AdvDM and FSGM even in its worst case, the JPEG compression. However, the huge performance drop when compressed still signals the need to design more robust protection algorithms against image compression.

\begin{table}[h]
\vspace{-1mm}
\caption{The robustness of the defensive algorithms against four commonly seen data corruptions. The model finetuned is SD v1.5 and the text-encoder is frozen. We assume $\bm{c_{prot}} = \bm{c_{explo}}$ for this specific experiment.}
\label{table7}
\centering
\begin{adjustbox}{width=.45\textwidth}    
\begin{tabular}{c | c | c c c c }
\toprule
Corruption & Data & FDS $(\downarrow)$ & FID $(\uparrow)$ & IQS $(\downarrow)$ & BRISQUE $(\uparrow)$ \\
\midrule
\multirow{4}{*}{Cropping} & AdvDM & 0.379 & 258.085 & -1.460 & 22.655 \\
& FSGM & 0.376 & 255.739 & -0.573 & 18.926 \\
& ASPL & 0.369 & 268.892 &  -1.850 & 22.319 \\
& PID  & \textbf{0.246} & \textbf{275.468} & \textbf{-6.290} & \textbf{24.183} \\
\midrule
\multirow{4}{*}{Smoothing} & AdvDM & 0.388 & 211.059 & \textbf{-3.351} & 18.013 \\
& FSGM & 0.388 & \textbf{229.721} & 1.391 & 16.403 \\
& ASPL & 0.377 & 223.193 & -2.210 & 17.212 \\
& PID  & \textbf{0.213} & 184.483 & 0.108 & \textbf{47.121} \\
\midrule
\multirow{4}{*}{Denoising} & AdvDM & 0.391 & 230.016 & \textbf{-1.656} & 20.457 \\
& FSGM & 0.396 & 230.049 & 2.108 & 17.639 \\
& ASPL & 0.372 & \textbf{248.910} & 0.326 & 21.292\\
& PID  & \textbf{0.213} & 184.483 & 0.108 & \textbf{42.440} \\
\midrule
\multirow{4}{*}{Compression} & AdvDM & 0.386 & 229.973 & -5.768 & 25.340 \\
& FSGM & 0.390 & 225.208 & -3.547 & 24.042\\
& ASPL & 0.354 & \textbf{267.039} & \textbf{-6.644} & \textbf{27.983} \\
& PID  & \textbf{0.345} & 221.601 & 0.287  & 20.510 \\
\bottomrule
\end{tabular}%
\end{adjustbox}%
\end{table} %

\section{Conclusion}
\par In this paper, we delve into the reliability of current data protection algorithms against LDMs without the prompt-consistency assumption. Our investigation reveals that the prompt-related defenses could suffer notable performance decreases when the data exploiters intentionally craft fine-tuning prompts. Motivated by the visual encoder's independence from the textual prompts, we thoroughly analyze how perturbing the visual encoder impacts the fine-tuning process and propose a prompt-independent defense algorithm named PID. With the empirically validated effectiveness of PID and its ability to enhance existing algorithms, we believe that our proposed prompt-independent algorithm marks an important step toward reliable protection of data from exploitation by LDMs.

\section*{Acknowledgements}
Yisen Wang was supported by National Key R\&D Program of China (2022ZD0160300), National Natural Science Foundation of China (62376010, 92370129), Beijing Nova Program (20230484344), and CCF-Baichuan-EB Fund.

\section*{Impact Statement}
\par Marching towards the overarching goal of privacy-conscious artificial intelligence, particularly in the realm of privacy-aware deep image generative models, our research offers significant empirical and technical advancements to this critical domain. With our work known by the broader civilians, we are confident that our work will bolster public trust in artificial intelligence, further mitigating the risks posed to personal privacy by generative models, and ultimately building trustworthy AI. 

\bibliography{example_paper}
\bibliographystyle{icml2024}

\appendix
\onecolumn
\section{Experimental Details}
\label{Appendix A}
\subsection{Metric Definition}
\par Throughout our paper we evaluate the fine-tuning results using four matrics, covering different properties of generated images. We define the \textit{Face Detection Score (FDS)} as the average cosine similarity between the embeddings of training images and the embeddings of generated images, where the embedding are given by a MTCNN \cite{mtcnn} pre-trained on a large-scale facial dataset VGGFace2 \cite{cao2018vggface2}. FDS mainly captures the semantic similarity between the images and evaluates whether the facial information is learned. Fréchet Inception Distance (FID) \cite{FID} is a metric evaluating the distance between the distribution of the generated images and the distribution of training images, evaluating the model's master of the image from another perspective. The \textit{Image Quality Score (IQS)} is defined to assess the quality of the generated images with the powerful visual-language model CLIP \cite{clip}. Concretely, we compute the cosine similarity between the clip embedding of the generated images and two sentences, \{\textit{A high-quality photo, A low-quality photo}\} respectively. We report a $10^3$ scale of the average differences between the two cosine similarities. Finally, we also adopt Blind/Referenceless Image Spatial Quality Evaluator (BRISQUE)  \cite{brisque}, a no-reference image quality assessment metric, to evaluate the image quality. For each LDM, we generate $20$ images with the 3 prompts respectively \{A photo of sks person, A selfie of sks person, A DSLR portrait of sks person\} and report averaged results over the $60$ generated images. When generating data from the diffusion models, we use 50-step DDPM sampling and set the negative prompt to be \textit{noisy, low quality, artifacts, poorly drawn face}.
\par The intuition behind us selecting the $c_{explo}$ resulting in the highest FDS to represent the performance in the prompt mismatch case is that higher FDS means the generated images have higher facial similarity with the training data, which is usually a sign for good fine-tuning result. 

\subsection{Data} 
\par We mainly adopt the CelebA-HQ \cite{celeba} dataset in our experiments. The dataset consists of thousands of celebrities, with approximately 10 photos each, and the image size is 512x512. We randomly draw 10 celebrities from the dataset and choose 4 images out of every set of images. Normally, $4 \sim 5$ images are sufficient for the LDMs to grasp the core concept in the images \cite{dreambooth, textual_inversion}. For experiments on VGGFACE \cite{cao2018vggface2}, we also randomly draw 10 instances from the dataset and choose 4 images out of every set of images.

\subsection{Fine-tuning Hyper-parameter}
\par We fine-tune the Stable Diffusion \cite{stablediffusion} models with DreamBooth \cite{dreambooth} and LoRA. In Table \ref{table8}, we list the specific fine-tuning details. 

\begin{table}[h]
\centering
\caption{Fine-tuning hyper-parameters.}
\label{table8}
\begin{subtable}[t]{1.0\textwidth}
\centering
\begin{adjustbox}{width=0.6\textwidth}
\begin{tabular}{ c | c | c | c | c | c | c}
\toprule
     Version & Freeze-TE & LR & Steps & Batch 
     Size & Grad. Accu. & Output Res. \\
\midrule
    1.5 & Yes & 2e-6 & 1000 & 1 & 1 & 512x512 \\
    1.5 & No  & 2e-6 & 500  & 1 & 1 & 512x512 \\
    2.1 & Yes & 1e-5 & 1000 & 1 & 1 & 728x728 \\
    2.1 & No  & 2e-6 & 500  & 1 & 1 & 728x728 \\
\bottomrule
\end{tabular}
\end{adjustbox}
\subcaption{Fine-tuning hyper-parameters for Dreambooth on CelebA-HQ.}
\end{subtable}

\begin{subtable}[t]{1.0\textwidth}
\centering
\begin{adjustbox}{width=0.6\textwidth}
\begin{tabular}{ c | c | c | c | c | c | c}
\toprule
     Version & Freeze-TE & LR & Steps & Batch 
     Size & Grad. Accu. & Output Res. \\
\midrule
    1.5 & Yes & 2e-6 & 1200 & 1 & 1 & 512x512 \\
    1.5 & No  & 2e-6 & 1000  & 1 & 1 & 512x512 \\
    2.1 & Yes & 1e-5 & 1200 & 1 & 1 & 728x728 \\
    2.1 & No  & 2e-6 & 1000  & 1 & 1 & 728x728 \\
\bottomrule
\end{tabular}
\end{adjustbox}
\subcaption{Fine-tuning hyper-parameters for Dreambooth on VGGFACE.}
\end{subtable}

\begin{subtable}[t]{1.0\textwidth}
\centering
\begin{adjustbox}{width=0.6\textwidth}
\begin{tabular}{ c | c | c | c | c | c | c}
\toprule
     Version & Rank & LR & Steps & Batch 
     Size & Grad. Accu. & Output Res. \\
\midrule
    1.5 & 16  & 1e-4 & 800  & 2 & 1 & 512x512 \\
    2.1 & 32  & 1e-4 & 1600  & 2 & 1 & 728x728 \\
\bottomrule
\end{tabular}
\end{adjustbox}
\subcaption{Fine-tuning hyper-parameters for LoRA on CelebA-HQ. We apply the low-rank adapters to both the UNet and the text-encoder.}
\end{subtable}
\end{table}

\subsection{Defense Hyper-parameter}
\par We mainly consider three defense algorithms in this paper AdvDM \cite{advDM}, FSGM, and ASPL \cite{antidreambooth}. For AdvDM \cite{advDM}, we implement the protection loss as $L_{semantic} + \lambda L_{textural}$ with $\lambda=0.05$, where the $L_{semantic}$ is defined as the training loss of the Stable Diffusion and the textural loss is defined as the $\ell_2$ distance between the mean value of the latent distributions. We run the algorithm for $100$ steps (the default number of steps is 40 in the paper) with step size equaling $\epsilon/10$. For FSGM, we run the defense for 100 iterations, with step size being $\epsilon/10$ and 1-step gradient accumulation. For ASPL, we run the defense for 50 iterations, for each iteration, the surrogate model is updated for 3 steps and the perturbations are updated for 6 steps. The step size is also set to be $\epsilon/10$. We set the reference images for training the surrogate models of ASPL and FSGM the same as the images to be protected. Note that we by no means intentionally run the algorithms for fewer steps to boost our proposed defense. The steps are recommended by the original as the default setting and running more steps will not improve their performance significantly. In all, for fair comparison and time consideration, we adopt their default configurations when it comes to the iterations to run.

\subsection{Fine-tuning Prompts when $\bm{c_{prot}} \neq \bm{c_{explo}}$}
\par The $\bm{c_{prot}}$ is fixed to be \textit{A photo of sks person}. We consider 3 $\bm{c_{explo}}$ when $\bm{c_{prot}} \neq \bm{c_{explo}}$ and we encourage future works to try on more diverse prompts or even using soft prompts that are optimized for specific goals for fine-tuning. The exhaustive list of which is 
\begin{itemize}
    \item \textit{A photo of sks person.}
    \item \textit{A photo of sks face.}
    \item \textit{A DSLR portrait of sks person. }
\end{itemize}

\section{More Results for The Prompt-Mismatch Scnenario}
\label{AppendixB}
\subsection{Qualitive Results}
\par
\begin{figure}[t]
\centering
    \begin{subfigure}[t]{1.0\textwidth}
    \centering
    \includegraphics[width=0.86\textwidth]{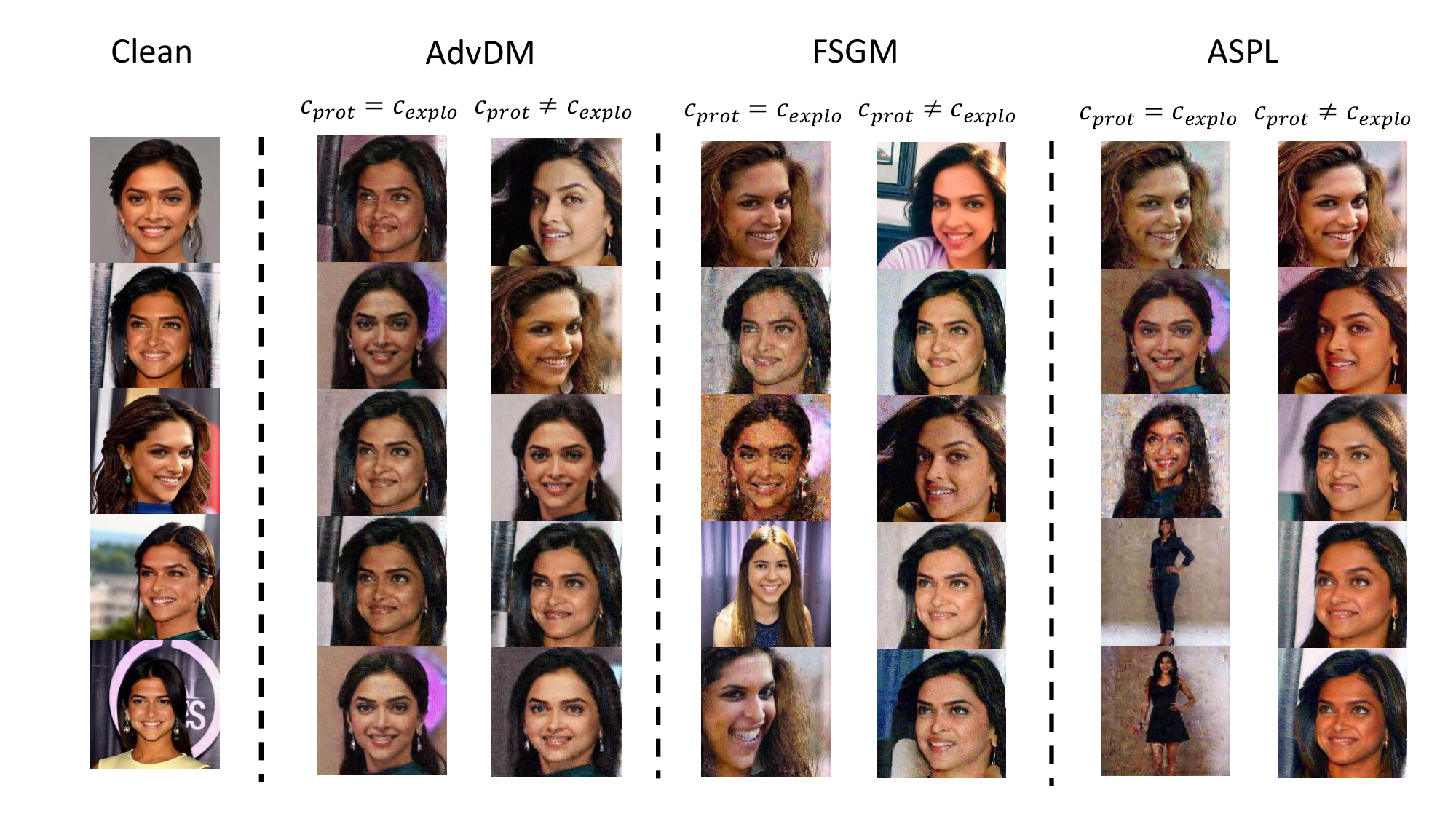}
    \caption{Images generated by SD v1.5 fine-tuned on clean/protected data with the text encoder frozen. The instance is from CelebA-HQ.}
    \label{figure 7-1}
    \end{subfigure}
    
    \begin{subfigure}[t]{1.0\textwidth}
    \centering
        \includegraphics[width=0.86\textwidth]{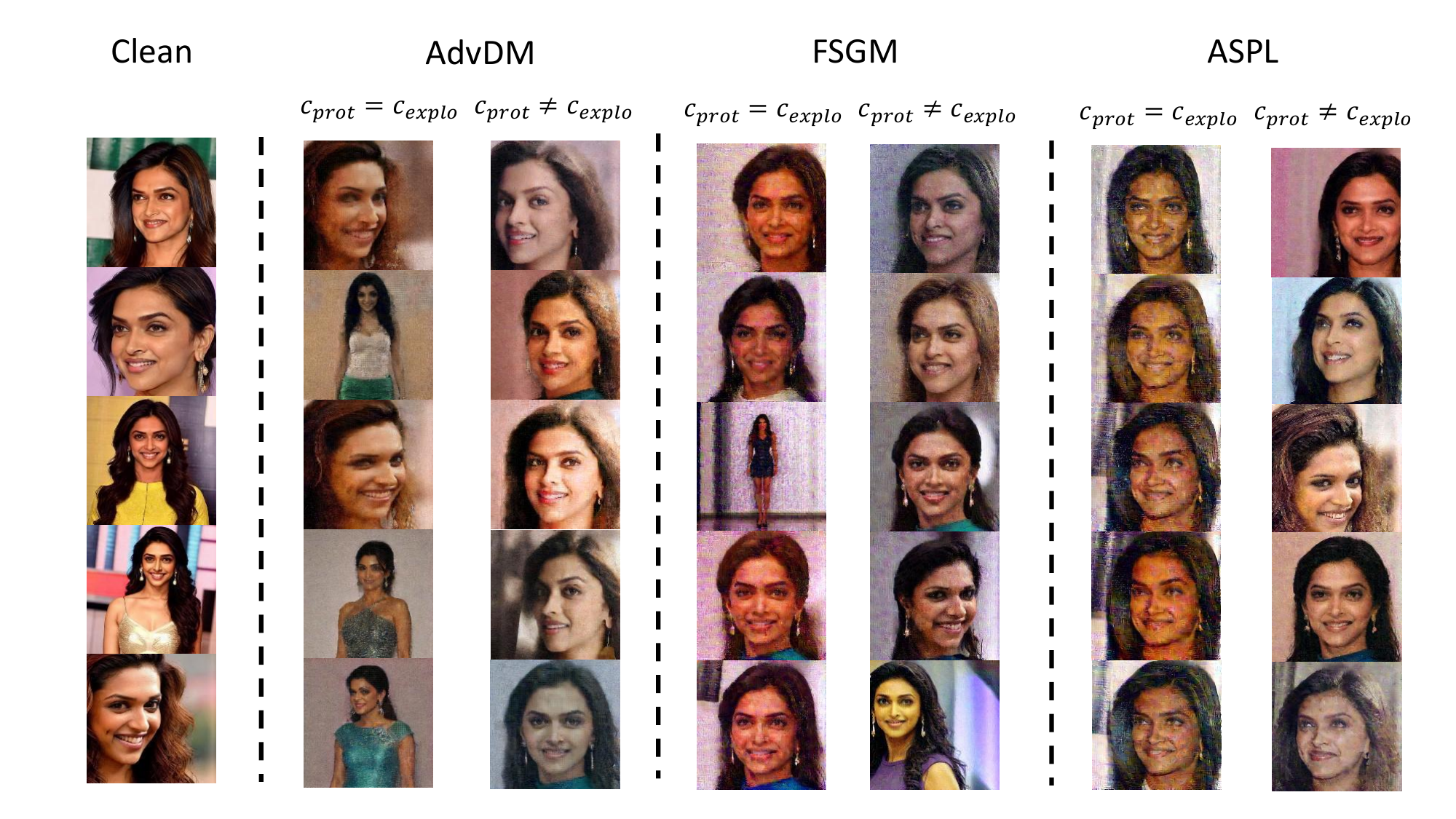}
    \caption{Images generated by SD v1.5 fine-tuned on clean/protected data with the text encoder unfrozen. The instance is from CelebA-HQ.}
    \label{figure 7-2}
    \end{subfigure}
    \caption{Influence of prompt-mismatch, i.e., $\bm{c_{prot}}\neq\bm{c_{explo}}$, on the protective performance.}
\end{figure}

 To better illustrate the performance degradation of the current defense algorithm under the prompt-mismatch scenario, i.e. $\bm{c_{prot}} \neq \bm{c_{explo}}$, we randomly select an instance from the CelebA-HQ\cite{celeba} dataset and protect the images with three symbolic defense algorithms, AdvDM \cite{advDM}, FSGM \cite{antidreambooth} and ASPL \cite{antidreambooth}. For each defense, we display 5 images generated by the Stable Diffusion v1.5 (SD v1.5) model fine-tuned on the data protected by it. We display the case where the text-encoder is frozen and unfrozen during fine-tuning in Figure \ref{figure 7-1} and \ref{figure 7-2} respectively.  

\subsection{Discussion on the Prompt-dependent Effect}
\par The intuition behind the prompt dependency lies in the training loss of the LDM, which requires a textual condition $\bm{c_{prot}}$ as one of its parameters. Existing attacks all involve the training loss as part of the defense objective to optimize, which makes the resulting perturbations inevitably related to the condition $\bm{c_{prot}}$. 
\par A very straightforward way to break the correlation between the conditions and the perturbations is to aggregate across several prompts during optimization. However, aggregating through $k$ prompts increases the computational cost by $k$ times and makes the objective even harder to solve. In Table 
\ref{table10}, we show that a naive ensembling of prompts to generate the perturbations 
won't fundamentally resolve the issue. 
\par
\begin{figure*}[h]
\centering
    \begin{subfigure}[ht]{0.49\textwidth}
    \includegraphics[width=1.0\textwidth]
    {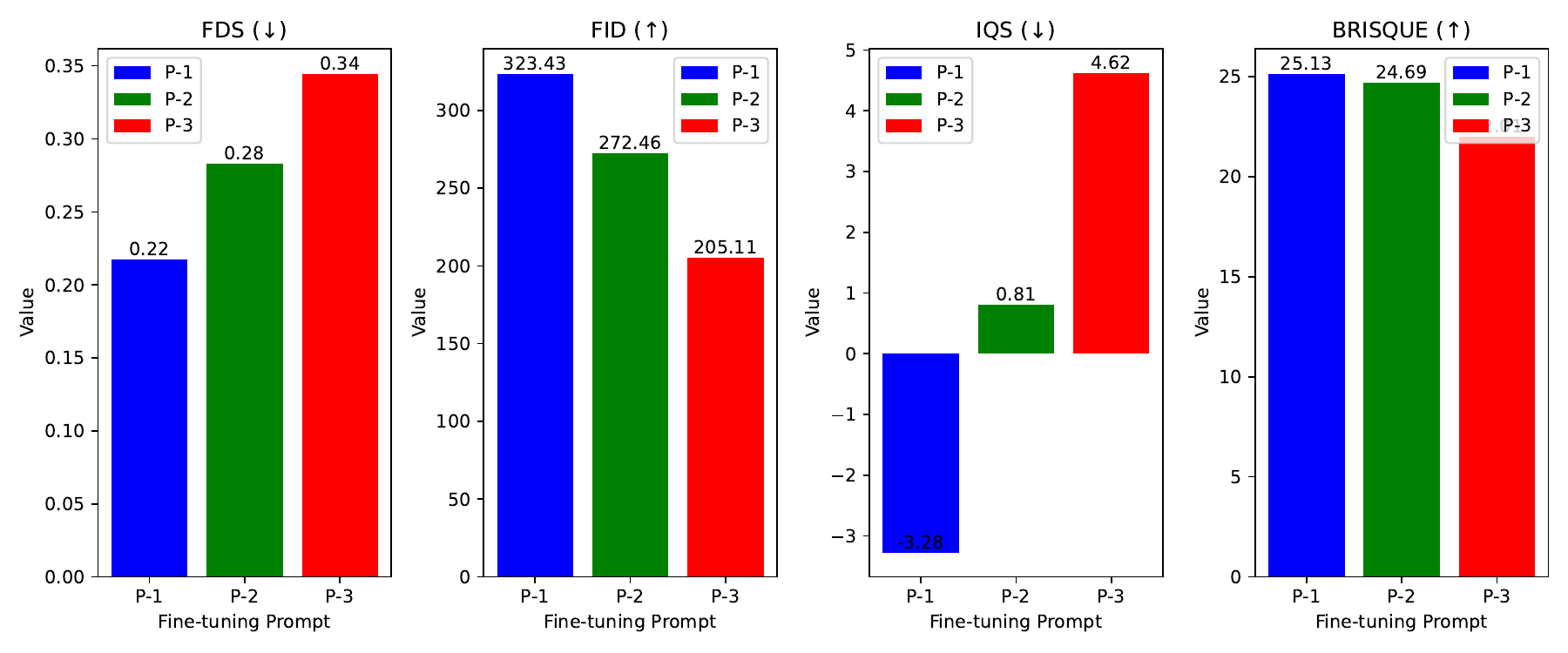}
    \caption{A type example of the prompt dependency. In the case where $\bm{c_{explo}} = \bm{c_{prot}}$, the protection performance is strong, producing low FDS and IQS. However, when $\bm{c_{explo}}$ is switched to P-2 and P-3, the fine-tuned model grasps the human face much better, leading to reasonable FDS and even positive IQS. The instance is from the VGGFACE dataset and is protected by FSGM.}
    \label{figure8-1}
    \end{subfigure} \hspace{2mm}%
    \begin{subfigure}[ht]{0.49\textwidth}
    \includegraphics[width=1.0\textwidth]{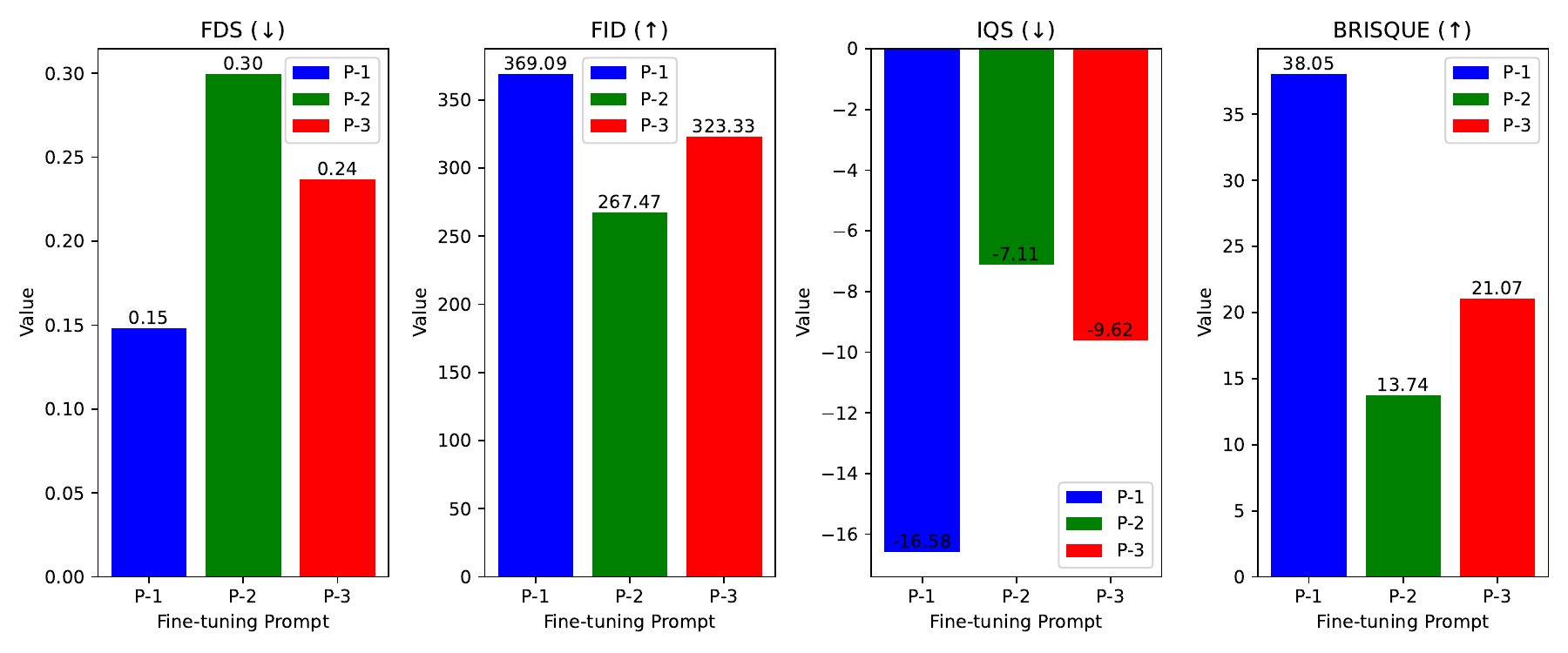}
    \caption{A typical example of the prompt dependency. In the case where $\bm{c_{explo}} = \bm{c_{prot}}$, the protection performance is fairly good. The resulting model generates images of very poor quality, evidenced by very low IQS and high BRISQUE. However, when $\bm{c_{explo}}$ is switched to P-2 and P-3, the fine-tuned model learns the human face notably better, leading to acceptable FDS and nearly normal BRISQUE. The instance is from the VGGFACE dataset and is protected by ASPL. }
    \label{figure8-2}
    \end{subfigure} \vspace{3mm}%
    \begin{subfigure}[ht]{0.49\textwidth}
    \includegraphics[width=1.0\textwidth]{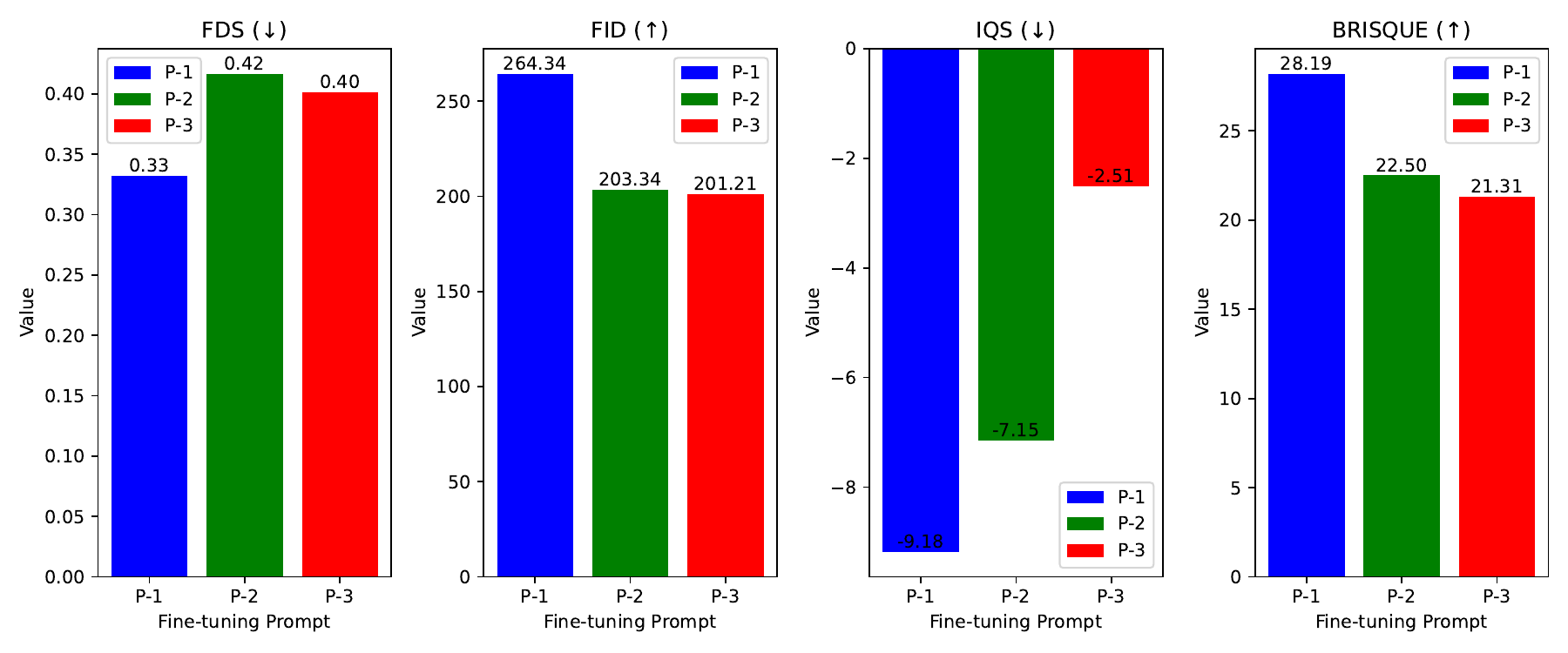}
    \caption{An example where the perturbations exhibit medium-level dependency on prompts. The protection is slightly weaker for P-2 and P-3. The instance is from the VGGFACE dataset and is protected by ASPL. }
    \label{figure8-3}
    \end{subfigure}\hspace{2mm} %
    \begin{subfigure}[ht]{0.49\textwidth}
    \includegraphics[width=1.0\textwidth]{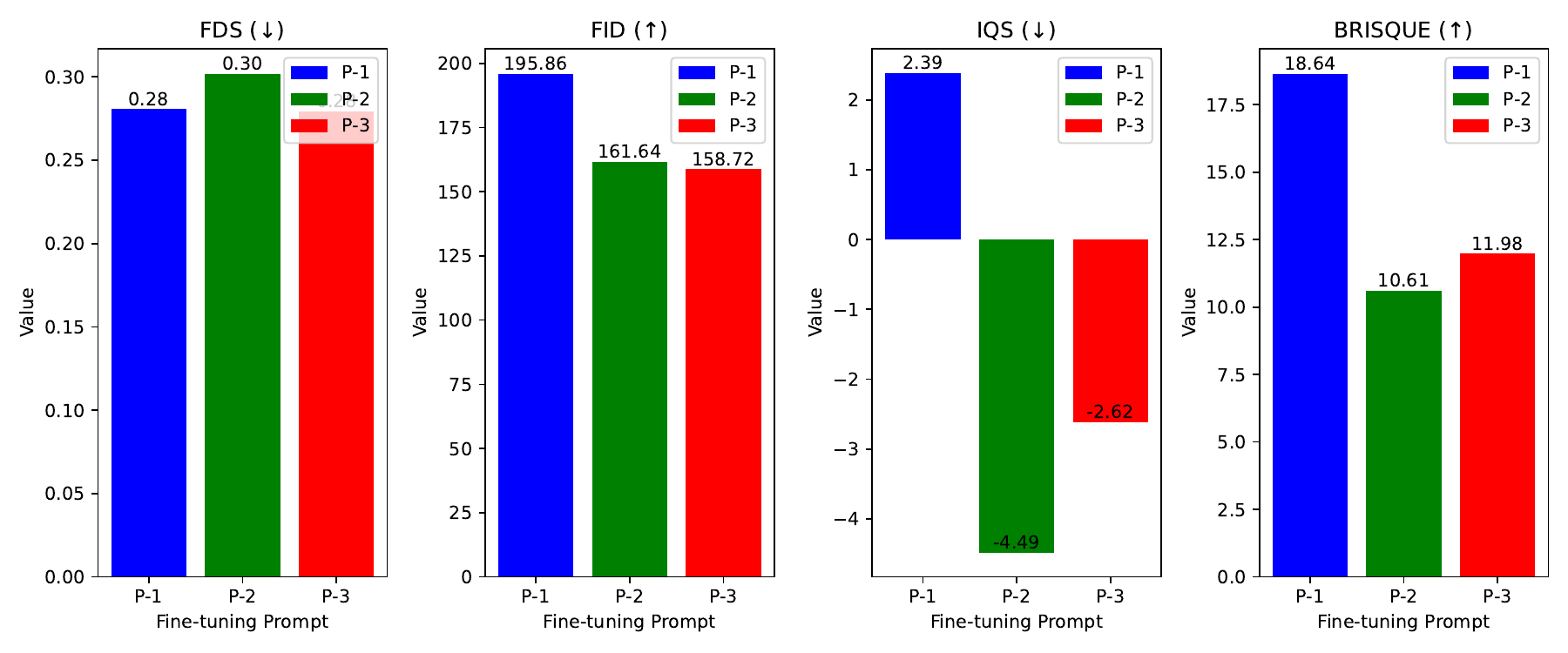}
    \caption{An example where the IQS shows counter-intuitive results, where the $\bm{c_{prot}}=\bm{c_{explo}}$ case exhibits best fine-tuning results. Even though, the dependency effect is still valid across the other three metrics. We by no means claim that $\bm{c_{prot}}\neq \bm{c_{explo}}$ always leads to a significant performance drop on every instance protected by any prompt-dependent defense. As shown in our quantitative results, our claim should be interpreted as an overall trend rather than a definite rule. The instance is from the VGGFACE dataset and is protected by FSGM. }
    \label{figure8-4}
    \end{subfigure} %

    \caption{instances selected to illustrate the effect of the mismatch between $\bm{c_{prot}}$ and $\bm{c_{explo}}$. P-1, P-2, and P-3 denote three $\bm{c_{explo}}$ defined in \ref{Appendix A} and $\bm{c_{prot}}$ is still \textit{A photo of sks person}. Note that these sub-figures serve as case studies rather than rigorous evidence for our claims. The fine-tuned model is SD v1.5 with text-encoder unfrozen.}
    \label{Figure8-Total}
\end{figure*}

\begin{figure*}[h]
\centering
    \begin{subfigure}[ht]{0.49\textwidth}
    \includegraphics[width=1.0\textwidth]{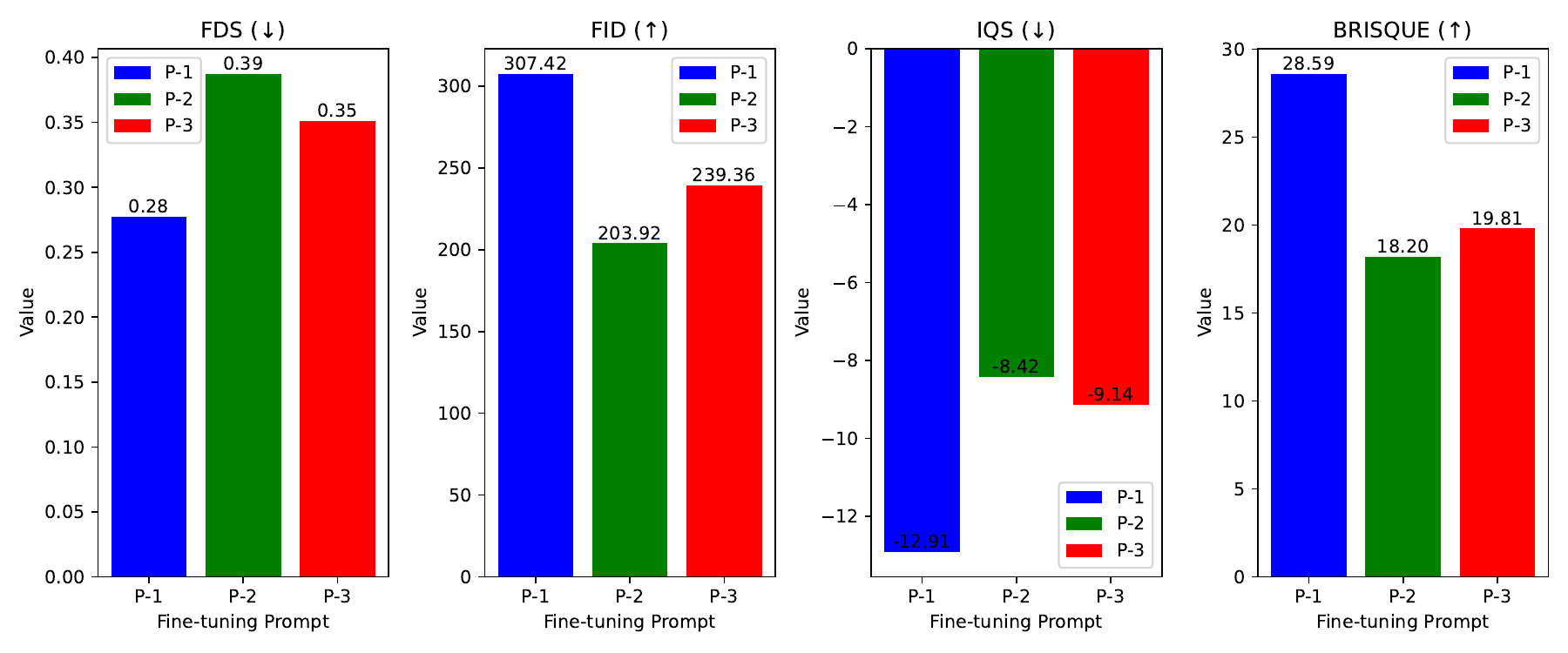}
    \caption{Prompt-dependent defense: FSGM \cite{antidreambooth}}
    \label{Figure9-1}
    \end{subfigure} \hspace{1mm}%
    \begin{subfigure}[ht]{0.49\textwidth}
    \includegraphics[width=1.0\textwidth]{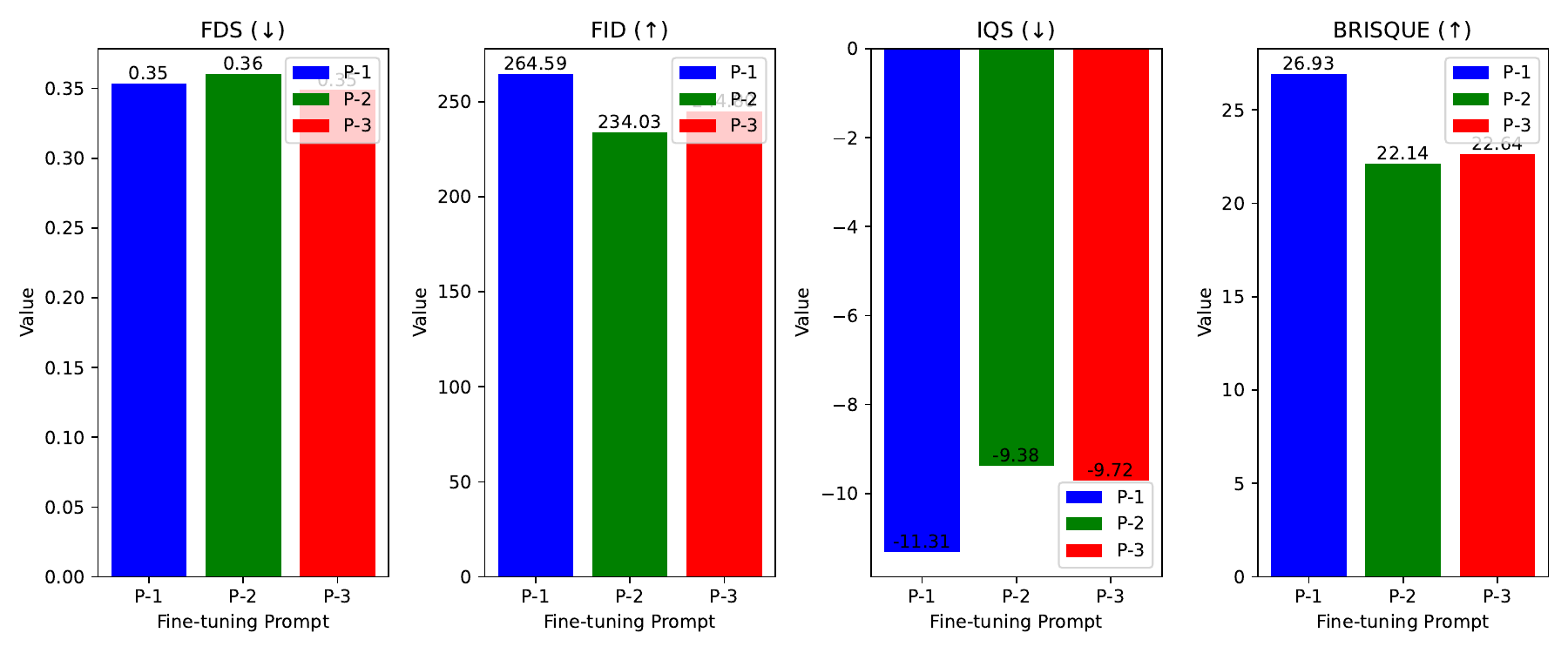}
    \caption{Prompt-dependent defense: APSL \cite{antidreambooth}}
    \label{Figure9-2}
    \end{subfigure} %
    \begin{subfigure}[ht]{0.49\textwidth}
    \includegraphics[width=1.0\textwidth]{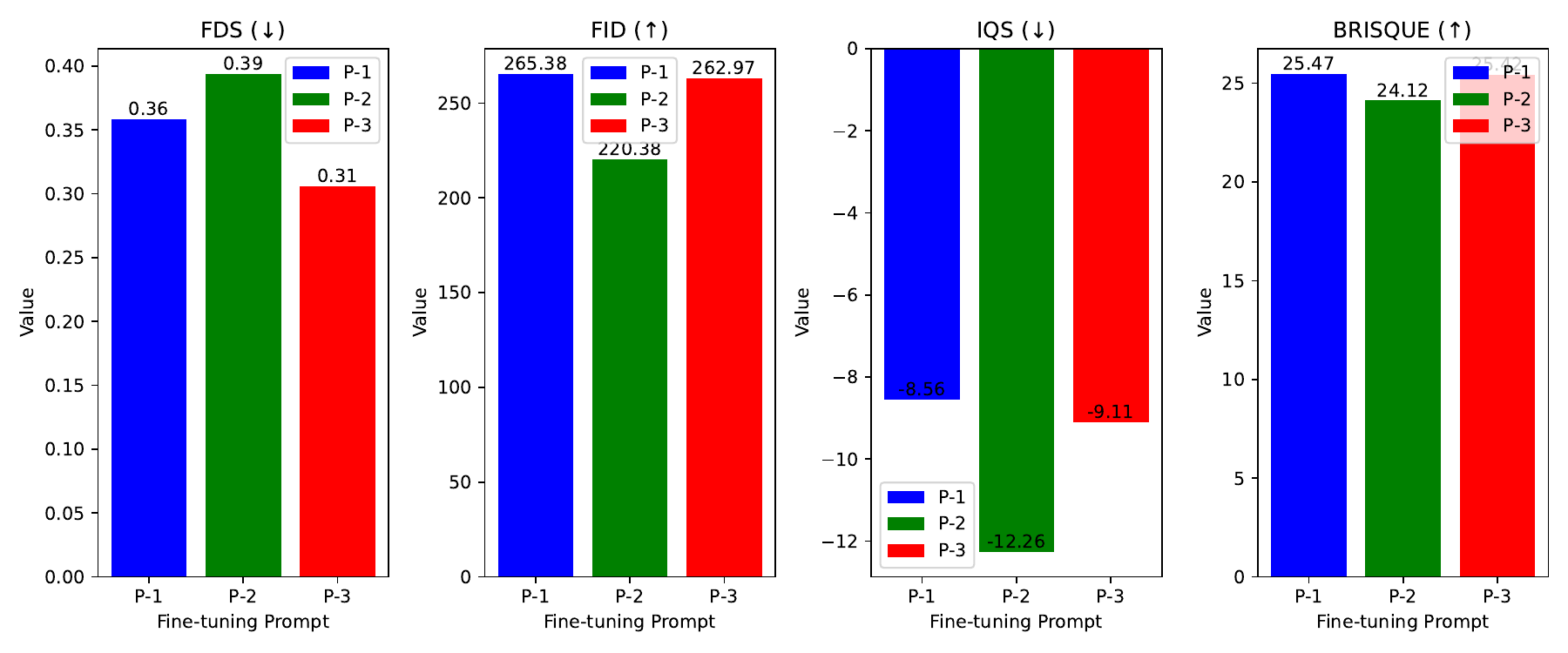}
    \caption{Prompt-dependent defense: AdvDM \cite{advDM}}
    \label{Figure9-3}
    \end{subfigure} \hspace{1mm}%
    \begin{subfigure}[ht]{0.49\textwidth}
    \includegraphics[width=1.0\textwidth]{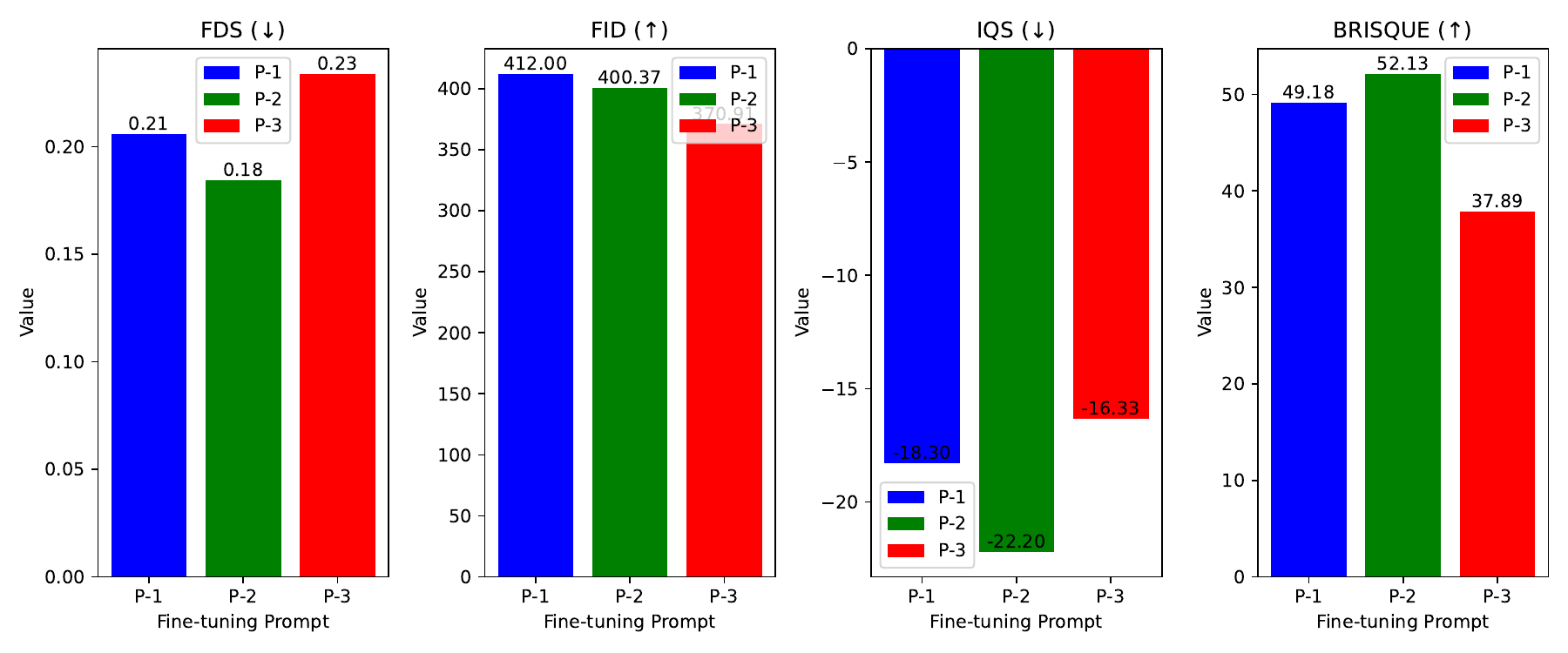}
    \caption{Prompt-dependent defense: PID}
    \label{Figure9-4}
    \end{subfigure}%
    
    \caption{Averaged fine-tuning results across the CelebA-HQ dataset for different $\bm{c_{explo}}$. For ASPL and FSGM, the defense performance in the $\bm{c_{prot}}=\bm{c_{explpo}}$ case out-performs the $\bm{c_{prot}}\neq \bm{c_{explo}}$ case, as evidenced by all four evaluation metrics. For AdvDM, which has a prompt-independent component, $L_{textural}$, as part of its defense target, better resilience against varied prompts is observed, as evident by P-3 achieving worse fine-tuning results than P-1. Contrary to the prompt-dependent defenses, the performance of PID remains steady when different fine-tuning prompts are used, which benefits from the perturbations' independence of the textual condition. The fine-tuned model is SD v1.5 with text-encoder unfrozen. }
    \label{Figure9-total}
\end{figure*}

 From Figure \ref{Figure8-Total} to Figure \ref{Figure9-total}, we provide more detailed results to discuss the effect of fine-tuning prompts on single instances.

\begin{table}[h]
\caption{Solving the prompt dependency with aggregation through several prompts during optimization. Here ASPL-k means we aggregate through k prompts. The fine-tuned model is Stable Diffusion v1.5 with text-encoder frozen and the dataset is CelebA-HQ. It can be seen that aggregation through more prompts doesn't bring significant improvement for the prompt-mismatch case. Since the exploiters theoretically have an infinite number of prompts to choose from, we cannot rely on such a method to achieve prompt-independent protection. }
\label{table9}
\centering
\begin{tabular}{c | c c c c c }
\toprule
& FDS $(\downarrow)$ & FID $(\uparrow)$ & IQS $(\downarrow)$ & BRISQUE $(\uparrow)$ \\
\midrule
ASPL-1 & \textbf{0.385} & 229.973 & -3.506 & 17.987 \\
ASPL-2 & 0.407 & 250.954 & -3.309 & 18.231 \\
ASPL-4 & 0.401 & \textbf{255.728} & \textbf{-4.064} & \textbf{19.596} \\
\bottomrule
\end{tabular}
\end{table}

\subsection{JPEG Results}
\par
\begin{figure}[t]
\centering
    \begin{subfigure}[t]{1.0\textwidth}
    \centering
    \includegraphics[width=0.86\textwidth]{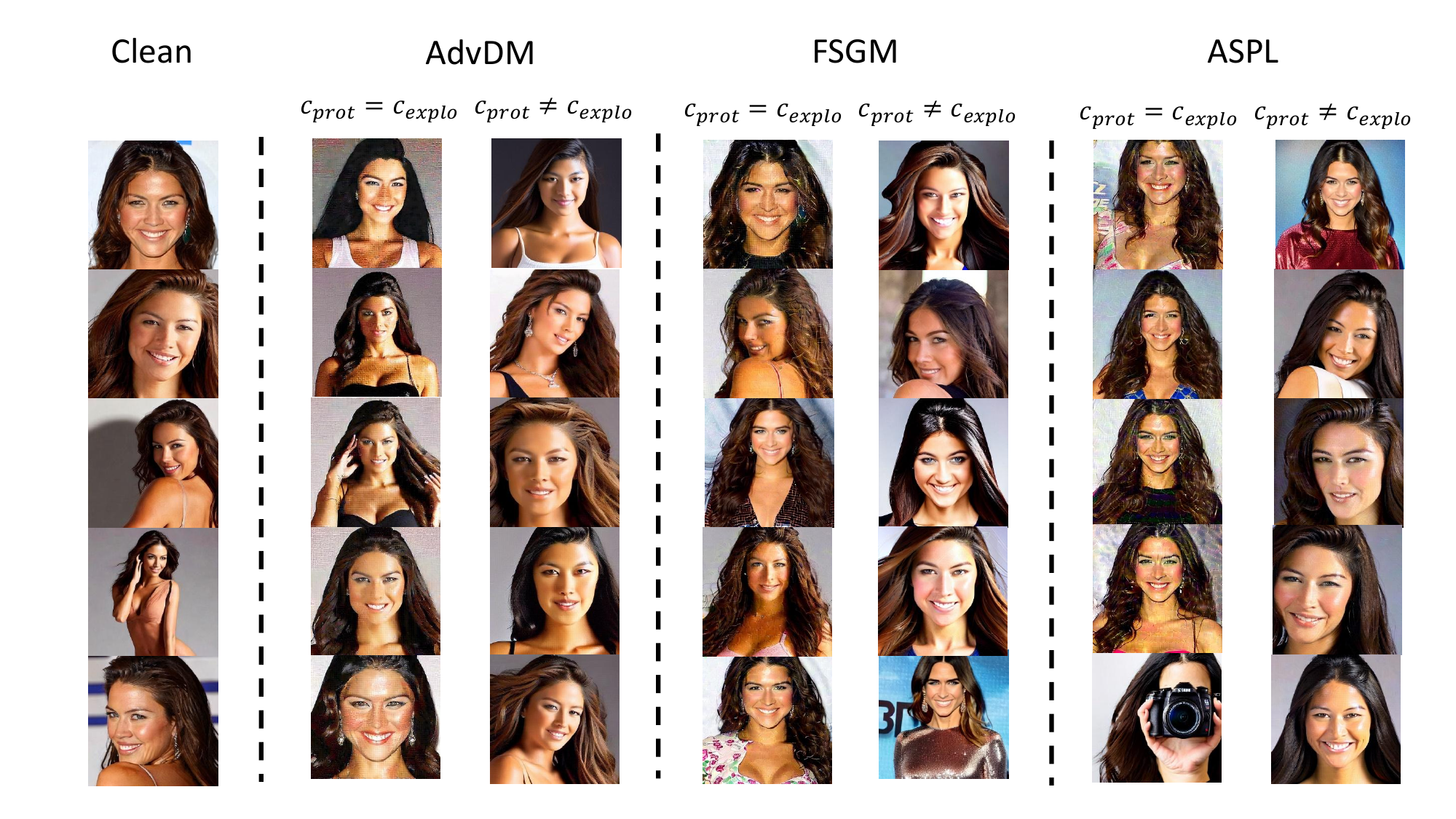}
     \caption{Images generated by SD v1.5 fine-tuned on clean/protected data with the text encoder frozen. The instance is from CelebA-HQ.}
    \label{figure 10-1}
    \end{subfigure}
    
    \begin{subfigure}[t]{1.0\textwidth}
    \centering
        \includegraphics[width=0.86\textwidth]{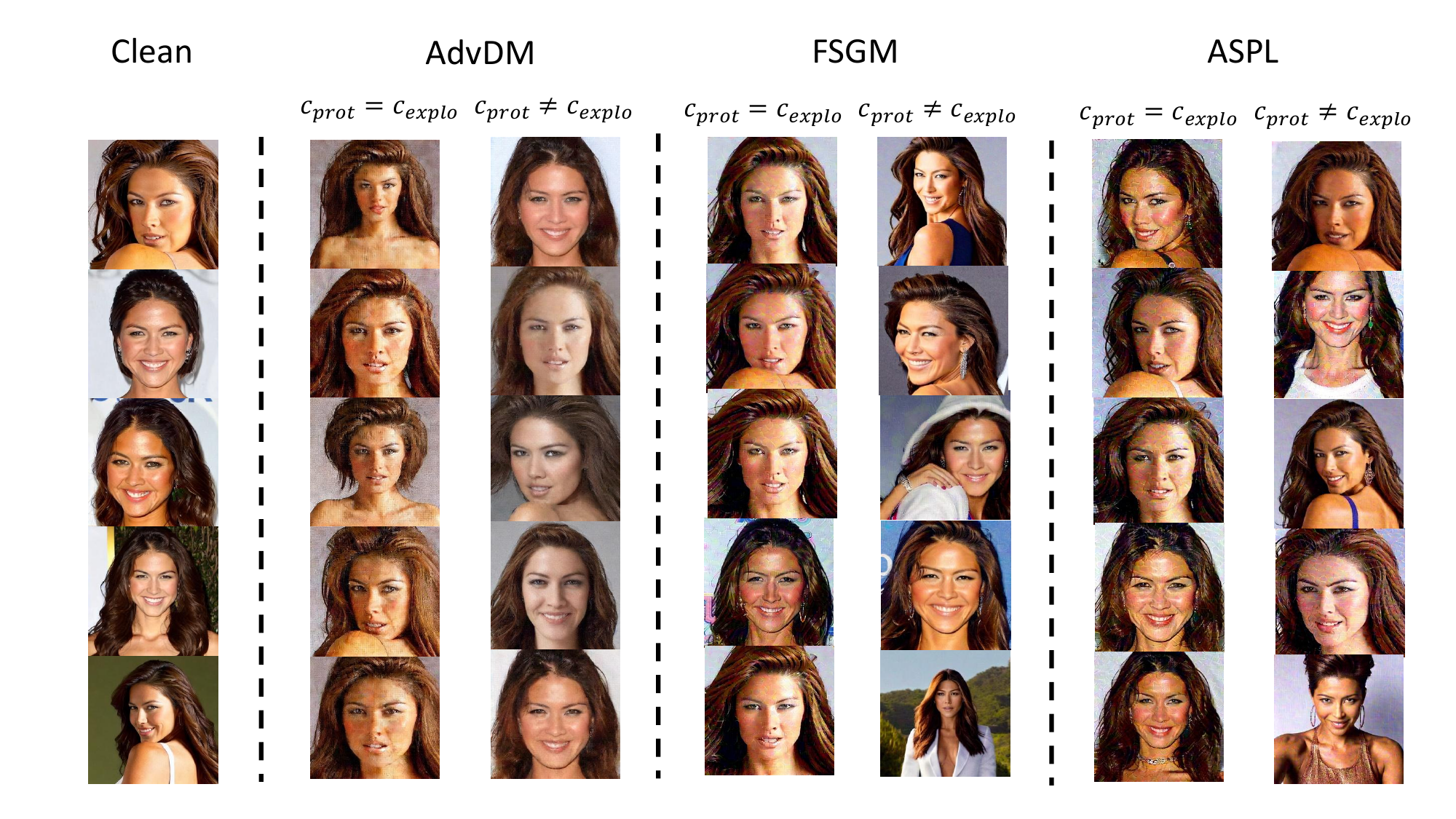}
     \caption{Images generated by SD v1.5 fine-tuned on clean/protected data with the text encoder unfrozen. The instance is from CelebA-HQ.}
    \label{figure 10-2}
    \end{subfigure}
    \caption{Influence of prompt-mismatch, i.e., $\bm{c_{prot}}\neq\bm{c_{explo}}$, on the protective performance. Protected images are saved in JPEG format.}
\end{figure}

 JPEG compression has long been known to have an undiminishable effect on the perturbations added to images \cite{JPEG}. We show quantitative and qualitative results for the cases where the protected images are saved in JPEG format in Table \ref{table10}, Figure \ref{figure 10-1}, and Figure \ref{figure 10-2} respectively.

\begin{table}[h]
\centering
\caption{ The quantitative results showing the performance of the prompt dependent defenses in cases where textual prompts are matched ($\bm{c_{prot}}=\bm{c_{explo}}$) and mismatched ($\bm{c_{prot}} \neq \bm{c_{explo}}$) between the protection stage and the exploration stage. \textbf{Freeze-TE} means whether we Freeze-the-Text-Encoder during fine-tuning. The arrows symbolize the direction of better protection. Perturbed images are saved in JPEG format in this table.}
\label{table10}
\begin{adjustbox}{width=.45\textwidth}    
\begin{tabular}{c | c | c c c c }
\toprule
Freeze-TE & Data & FDS $(\downarrow)$ & FID $(\uparrow)$ & IQS $(\downarrow)$ & BRISQUE $(\uparrow)$ \\
\midrule
\multirow{7}{*}{\cmark} & Clean & 0.480 & 144.570 & 4.130 & 14.757 \\
& AdvDM: $c = c_1$ & 0.386 & 229.973 & -5.768 & 25.340 \\
& AdvDM: $c \neq c_1$ & 0.421 & 157.750 & -4.962 & 20.102 \\
& FSGM: $c = c_1$ & 0.390 & 225.208 & -3.547 & 24.042 \\
& FSGM: $c \neq c_1$ & 0.424 & 211.879 & -1.771 & 16.650 \\
& ASPL: $c = c_1$ & 0.354 & 267.039 & -6.644 & 27.983 \\
& ASPL: $c \neq c_1$ & 0.385 & 229.973 & -3.506 & 17.987 \\
\midrule
\multirow{7}{*}{\xmark} & Clean & 0.557 & 128.870 & 7.104 & 18.445 \\
& AdvDM: $c = c_1$ & 0.275 & 199.338 & -4.866 & 25.723 \\
& AdvDM: $c \neq c_1$ & 0.377 & 182.473 & -4.286 & 17.857 \\
& FSGM: $c = c_1$ & 0.327 & 258.643 & -8.261 & 22.822 \\
& FSGM: $c \neq c_1$ & 0.412 & 206.979 & -3.316 & 21.238 \\
& ASPL: $c = c_1$ & 0.266 & 311.892 & -9.727 & 22.019 \\
& ASPL: $c \neq c_1$ & 0.340 & 254.840 & -7.588 & 20.326  \\
\bottomrule
\end{tabular}%
\end{adjustbox}%
\vspace{-1mm}
\end{table}

\section{Quantitive Results for Additional Datasets and Training Algorithm}
\label{AppendixC}

\subsection{LoRA Resutls}
\par We report the results of using LoRA \cite{lora} as the fine-tuning algorithm on the CelebA-HQ \cite{celeba} dataset in Table \ref{table11}. We apply the low-rank adapters to both the UNet \cite{unet} and the CLIP \cite{clip} text-encoder.
\begin{table}[h]
\centering
\caption{Evaluation of defense algorithms under the $\bm{c_{prot}} = \bm{c_{explo}}$ and $\bm{c_{prot}} \neq \bm{c_{explo}}$ scenarios when using LoRA \cite{lora} for fine-tuning. The best-performing defense under each metric is marked with \textbf{bold}.}
\label{table11}
\begin{subtable}[t]{0.95\textwidth}
\centering
\begin{adjustbox}{width=1.0\textwidth}
\begin{tabular}{ c | c | c  c c c c | c c c c }
\toprule
\multirow{2}{*}{Scenario}& \multirow{2}{*}{Data} & \multicolumn{4}{c}{Stable Diffusion v1.5} &&   \multicolumn{4}{c}{Stable Diffusion v2.1} \\
\cline{3-4} \cline{5-11}
& &  FDS$(\downarrow)$ & FID$(\uparrow)$ & IQS$(\downarrow)$ & BRISQUE$(\uparrow)$ && FDS$(\downarrow)$ & FID$(\uparrow)$ & IQS$(\downarrow)$ & BRISQUE$(\uparrow)$ \\
\midrule
     & Clean & 0.465 & 224.829 & 7.353 & 13.116 && 0.459 & 222.446 & 14.581 & 7.977 \\
     \midrule
        \multirow{4}{*}{$\bm{c_{prot}} = \bm{c_{explo}}$} 
        & AdvDM & 0.330 & 300.942 & -10.414 & 17.591 &&  0.219 & 354.107 & -3.662 & 35.402 \\
        & FSGM & 0.309 & \textbf{373.485} & -9.524 & 11.053 &&  0.173 & 423.777 & -12.004 & 41.896 \\
        & ASPL & 0.295 & 372.883 & -8.113 & 15.165 &&  0.174& 376.804 & -4.950 & 40.661 \\
        & PID  & \textbf{0.231} & 341.410 & \textbf{-18.782} & \textbf{46.977} && \textbf{0.163 }& \textbf{414.323} & \textbf{-17.694} & \textbf{59.272}\\
        
    \cmidrule{1-11}
    \multirow{4}{*}{$\bm{c_{prot}} \neq \bm{c_{explo}}$} 
        & AdvDM & 0.377 & 283.286 & -3.178 & 14.504 && 0.239 & 338.788 & -4.998 & 33.312 \\
        & FSGM  & 0.334 & 362.437 & -3.800 & 10.633 &&  0.240 & 391.125 & -8.771 & 34.302 \\
        & ASPL & 0.327 & \textbf{381.953} & -4.210 & 13.613 &&  0.225 & 352.907 & -4.033 & 38.890 \\
        & PID  & \textbf{0.276} & 322.451 & \textbf{-11.353} & \textbf{36.335} &&  \textbf{0.212} & \textbf{396.638} & \textbf{-18.327} & \textbf{50.168} \\
\bottomrule
\end{tabular}
\end{adjustbox}
\end{subtable}%
\end{table}

\subsection{VGGFACE Resutls}
\par We report the results for the VGGFACE dataset \cite{cao2018vggface2} in Table \ref{table12}.

\begin{table*}[h]
\centering
\caption{Quantitive evaluation of defense algorithms under the $\bm{c_{prot}} = \bm{c_{explo}}$ and $\bm{c_{prot}} \neq \bm{c_{explo}}$ scenarios on the VGGFACE dataset. The best-performing defense under each metric is marked with \textbf{bold}.}
\label{table12}
\begin{subtable}[t]{0.95\textwidth}
\centering
\begin{adjustbox}{width=1.0\textwidth}
\begin{tabular}{ c | c | c  c c c c | c c c c }
\toprule
\multirow{2}{*}{Scenario}& \multirow{2}{*}{Data} & \multicolumn{4}{c}{Frozen Text Encoder} &&   \multicolumn{4}{c}{Unfrozen Text Encoder} \\
\cline{3-4} \cline{5-11}
& &  FDS$(\downarrow)$ & FID$(\uparrow)$ & IQS$(\downarrow)$ & BRISQUE$(\uparrow)$ && FDS$(\downarrow)$ & FID$(\uparrow)$ & IQS$(\downarrow)$ & BRISQUE$(\uparrow)$ \\
\midrule
     & Clean & 0.442 & 164.203 & 16.359 & 12.290  && 0.512 & 134.21 & 17.63 & 7.37 \\
     \midrule
        \multirow{4}{*}{$\bm{c_{prot}} = \bm{c_{explo}}$} 
        & AdvDM & 0.332 & 211.935 & 0.928 & 13.853  &&  0.244 & 235.402 & -5.694 & 21.423 \\
        & FSGM & 0.294 & \textbf{282.948} & 0.064 & 19.111  && 0.235 & 285.889 & \textbf{-10.024} & 23.694 \\
        & ASPL &  0.312 & 274.449 & \textbf{-3.710} & 22.479 && 0.266 & 293.886 & -6.178 & 26.454 \\
        & PID  & \textbf{0.203} & 278.809 & -1.301 & \textbf{26.772}  &&  \textbf{0.223} & \textbf{295.264} & -4.730 & \textbf{27.273}  \\
        
    \cmidrule{1-11}
    \multirow{4}{*}{$\bm{c_{prot}} \neq \bm{c_{explo}}$} 
        & AdvDM  &  0.348 & 222.290 & -0.317 & 12.416  &&  0.329 & 214.461 & 1.853 & 16.445\\
        & FSGM  & 0.321 & 259.285 & -2.736 & 17.554 &&  0.329 & 232.522 & \textbf{-7.382} & 18.067 \\
        & ASPL  &  0.326 & \textbf{279.141} & \textbf{-2.767} & \textbf{21.150} &&  0.350 & 235.014 & -1.329 & 14.721\\
        & PID   &  \textbf{0.230} & 273.189 & -1.577 & 20.622  &&  \textbf{0.247} & \textbf{300.853} & -3.595 & \textbf{24.352}\\
\bottomrule
\end{tabular}
\end{adjustbox}
\subcaption{The fine-tuned model is Stable Diffusion v1.5.}
\end{subtable}%

\begin{subtable}[t]{0.95\textwidth}
\centering
\begin{adjustbox}{width=1.0\textwidth}
\begin{tabular}{ c | c | c  c c c c | c c c c }
\toprule
\multirow{2}{*}{Scenario}& \multirow{2}{*}{Data} & \multicolumn{4}{c}{Frozen Text Encoder} &&   \multicolumn{4}{c}{Unfrozen Text Encoder} \\
\cline{3-4} \cline{5-11}
& &  FDS$(\downarrow)$ & FID$(\uparrow)$ & IQS$(\downarrow)$ & BRISQUE$(\uparrow)$ && FDS$(\downarrow)$ & FID$(\uparrow)$ & IQS$(\downarrow)$ & BRISQUE$(\uparrow)$ \\
\midrule
     & Clean  &  0.455 & 227.395 & 15.580 & 10.413  && 0.482 & 200.339 & 13.935 & 7.524 \\
     \midrule
        \multirow{4}{*}{$\bm{c_{prot}} = \bm{c_{explo}}$} 
        & AdvDM  &  0.241 & \textbf{301.904} & 2.202 & 22.885  &&  0.321 & 252.942 & \textbf{-7.852} & 26.315\\
        & FSGM  & \textbf{0.236} & 275.736 & \textbf{-5.329} & \textbf{34.469}&&  \textbf{0.253} & \textbf{342.073} & -7.091 & 28.114 \\
        & ASPL & 0.253 & 280.514 & 0.827 & 27.108  &&  0.302 & 251.864 & -6.905 & \textbf{29.900}  \\
        & PID  &  0.301 & 282.013 & 1.749 & 15.406 && 0.321 & 267.520 & -1.258 & 27.993 \\
        
    \cmidrule{1-11}
    \multirow{4}{*}{$\bm{c_{prot}} \neq \bm{c_{explo}}$} 
        & AdvDM  &  \textbf{0.269} & \textbf{284.283} & -0.553 & 17.726 &&  0.364 & 219.191 & -1.257 & 24.797 \\
        & FSGM & 0.297 & 269.976 & \textbf{-9.208} & \textbf{32.260} && 0.351 & 226.134 & 1.560 & \textbf{29.667}  \\
        & ASPL &  0.296 & 240.470 & 1.612 & 25.668 &&  0.340 & 234.291 & \textbf{-5.408} & 27.554 \\
        & PID & 0.338 & 267.284 & 3.039 & 14.360  &&  \textbf{0.337} & \textbf{252.355} & 0.066 & 25.186 \\
\bottomrule
\end{tabular}
\end{adjustbox}
\subcaption{The fine-tuned model is Stable Diffusion v2.1.}
\end{subtable}%
\end{table*}

\subsection{$8/255$ Resutls}
\par We report the results of applying a tighter constraint on the perturbation budget, i.e. $\epsilon_\infty=8/255$ when running the defensive algorithms on the CelebA-HQ \cite{celeba} dataset in Table \ref{table13}. The fine-tuned model is Stable Diffusion v1.5.

\begin{table}[h]
\centering
\caption{Evaluation of defense algorithms under the $\bm{c_{prot}} = \bm{c_{explo}}$ and $\bm{c_{prot}} \neq \bm{c_{explo}}$ scenarios under the constraint that $\epsilon_\infty=8/255$. The fine-tuned model is SD v1.5. The best-performing defense under each metric is marked with \textbf{bold}.}
\label{table13}
\begin{subtable}[t]{1.0\textwidth}
\centering
\begin{adjustbox}{width=.6\textwidth}    
\begin{tabular}{c | c | c c c c }
\toprule
Freeze-TE & Data & FDS $(\downarrow)$ & FID $(\uparrow)$ & IQS $(\downarrow)$ & BRISQUE $(\uparrow)$ \\
\midrule
& Clean & 0.480 & 144.570 & 4.130 & 14.757 \\
\midrule
\multirow{4}{*}{\cmark} 
& AdvDM & 0.375 & 209.807 & -5.405 & 21.344 \\
& FSGM & 0.403 & 214.184 & -2.998 & 24.093 \\
& ASPL & 0.412 & 209.987 & -4.213 & 23.020\\
& PID & \textbf{0.256} & \textbf{248.038} & \textbf{-9.813} & \textbf{33.342}\\
\midrule
\multirow{4}{*}{\xmark} 
& AdvDM & 0.410 & \textbf{227.546} & -5.572 & 14.913 \\
& FSGM & 0.429 & 200.422 & -1.174 & 22.564 \\
& ASPL & 0.416 & 200.152 & -3.266 & 21.091 \\
& PID & \textbf{0.312} & 221.636 & \textbf{-9.949} & \textbf{27.391} \\
\bottomrule
\end{tabular}%
\end{adjustbox}%
\end{subtable}%
\end{table}

\section{Influence on Image Quality}
\label{AppendixD}
\par
\begin{figure}[th]
\centering
    \includegraphics[width=1.0\textwidth]{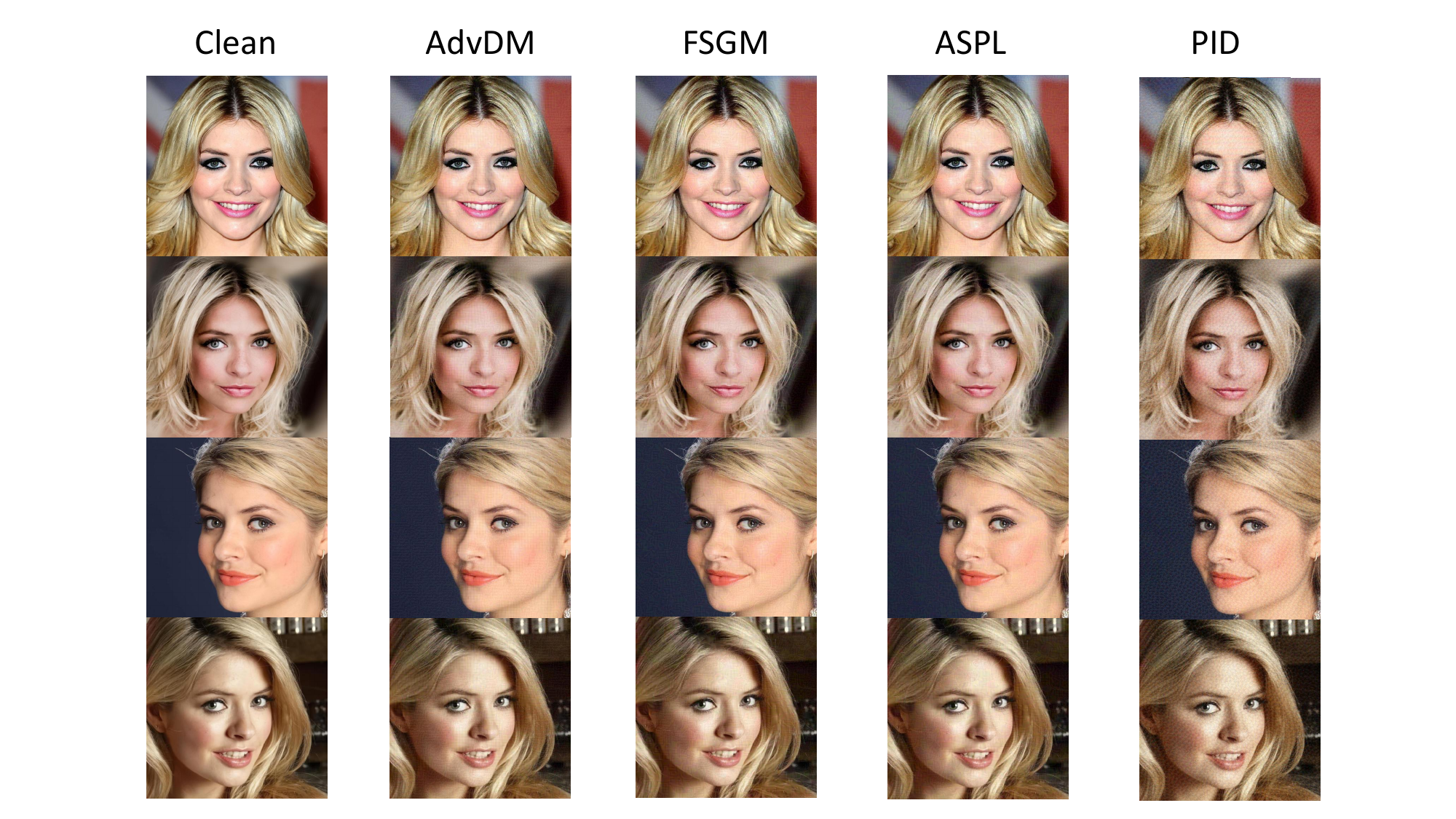}
    \caption{Comparision between the clean images and images protected by the four defensive algorithms. We use the SD v1.5 to generate these images. The pertubation budget is $\epsilon_\infty=8/255$.}
    \label{figure 11}
\end{figure}

 In this section, we assess the influence of PID on the image quality. We adopt three metrics to quantitively measure the influence. The three metrics are SSIM \cite{ssim}, PSNR, and LPIPS \cite{LPIPS} \footnote{We adopt the implementation from \hyperlink{piq}{https://github.com/photosynthesis-team/piq}}. We report the quantitive results in Table \ref{table14} and the qualitative comparisons are provided in Figure \ref{figure 11}. PID affects the images in a more aggressive way, which might contribute to its better robustness against image corruption and transferability across models.

\begin{table}[h]
\caption{The influence of defensive perturbations on image quality with three matrices measuring the difference between the perturbed images and the clean images. }
\label{table14}
\centering
\begin{tabular}{c | c c c }
\toprule
 & SSIM & PSNR & LPIPS \\
 \midrule
Random Noise & 0.82 & 30.90 & 0.20 \\
\midrule
AdvDM & 0.85 & 32.57 & 0.19 \\
FSGM & 0.91 & 35.76 & 0.19 \\
ASPL & 0.90 & 35.91 & 0.19 \\
PID  & 0.71 & 30.90 & 0.20 \\
\bottomrule
\end{tabular}
\end{table}

\section{Disscusion on PID and Image Editing}
\par Zero-shot image editing \cite{sdedit, kawar2023imagicediting} is another amazing ability of the LDMs and can also threaten civil privacy. Following the setting of \citet{photoguard}, we test the performance of PID in the image editing scenario. We protect the default image provided by \citet{photoguard} with the simple attack proposed by it, which targetedly manipulates the mean value and our defense. The model is the Stable Diffusion Inpainting \footnote{Downloaded from \hyperlink{SD-inpaint}{https://huggingface.co/runwayml/stable-diffusion-inpainting}} and both the attacks are obtained via PGD$_{100}$. The perturbation budget is $\epsilon_\infty=0.05$.
\begin{figure}[htbp]
\centering %
\begin{subfigure}[b]{0.35\textwidth}
    \centering
    \includegraphics[width=\linewidth]{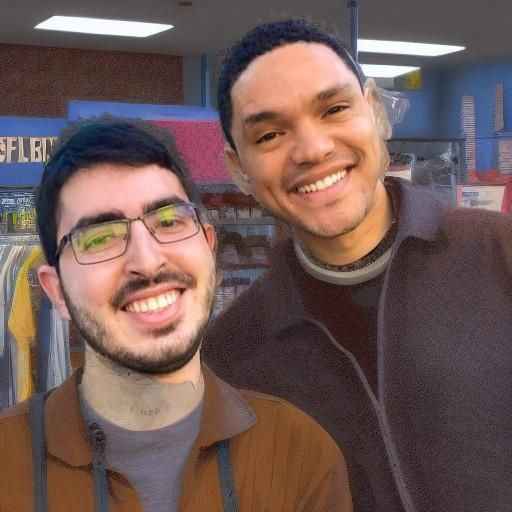}
    \caption{Simple attack}
    \label{fig12-2}
\end{subfigure}
~
\begin{subfigure}[b]{0.35\textwidth}
    \centering
    \includegraphics[width=\linewidth]{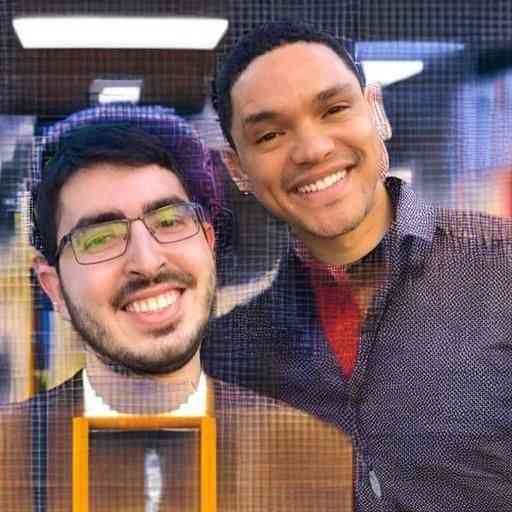}
    \caption{PID}
    \label{fig12-3}
\end{subfigure}
\caption{PID's performance in the image editing scenario following the setting of \citet{photoguard}. (a) the simple attack in \citet{photoguard}. (b) PID. The inference prompt is \textit{two men in a library}. }
\label{fig12}
\end{figure}
\par Observing Figure \ref{fig12}, the PID has a larger influence on the inpainting result than the simple attack, injecting meaningless noise and disrupting the image's semantics. However, we find that PID is yet not capable enough to adequately protect the images in this scenario, with the image still largely plausible. We leave the application of prompt-independent defense in the image inpainting scenario for future work. 

\clearpage
\section{More Visualization}
\label{AppendixE}
\par \textcolor{red}{\textbf{CAUTION: The images presented below may cause DISCOMFORT.}}

\begin{figure}[!b]
\label{figure 13}
\begin{subfigure}[b]{1.0\textwidth}
    \centering
    \includegraphics[width=0.95\textwidth, page=5]{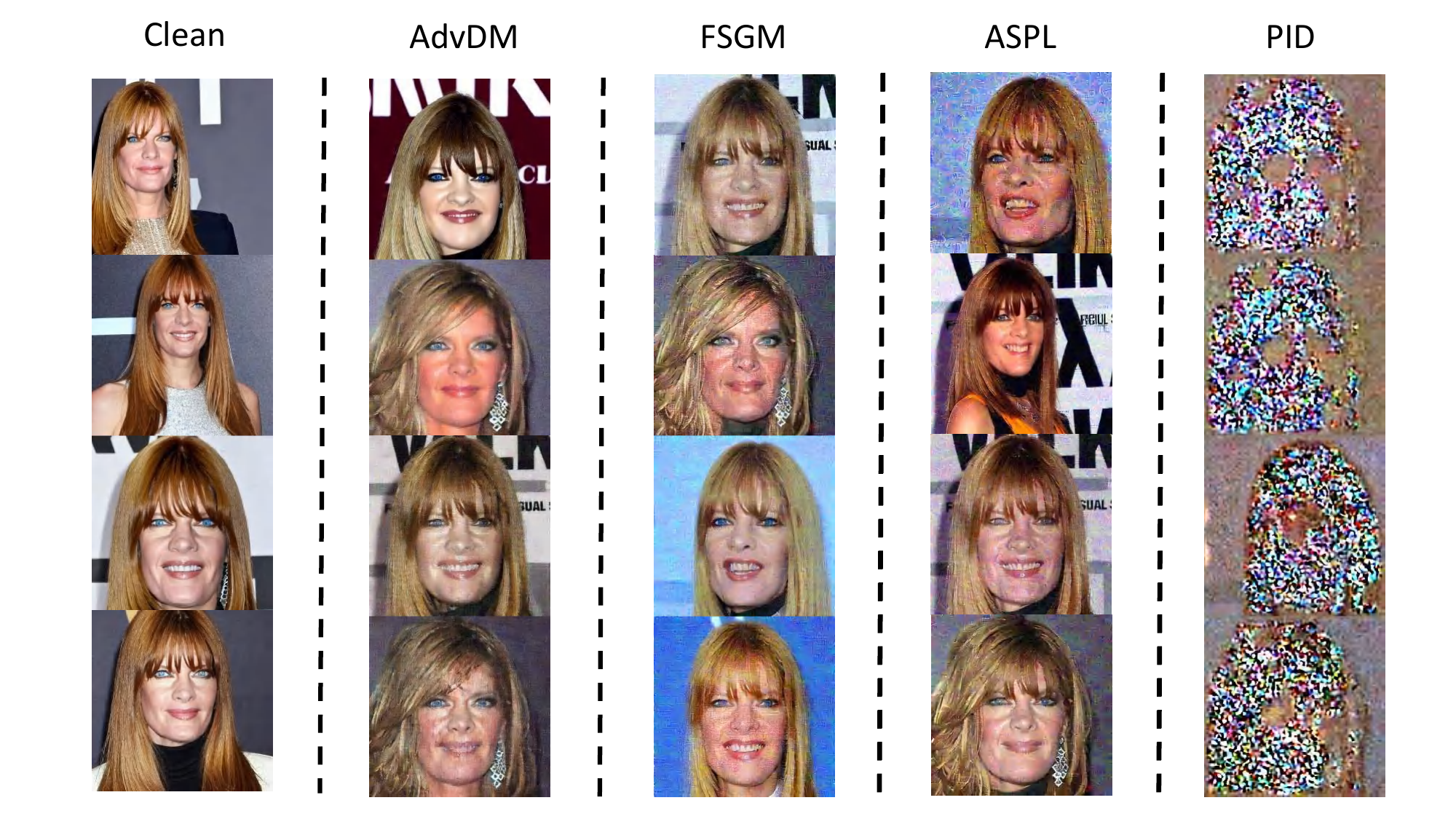}
\end{subfigure}%
\caption{Images synthesized by SD v1.5 fine-tuned on an instance from the CelebA-HQ protected by the defensive algorithms. The text-encoder is frozen during fine-tuning and the algorithm is Dreambooth. }
\end{figure}

\begin{figure}[h]
\label{figure 14}
\begin{subfigure}[b]{1.0\textwidth}
    \centering
    \includegraphics[width=0.95\textwidth, page=6]{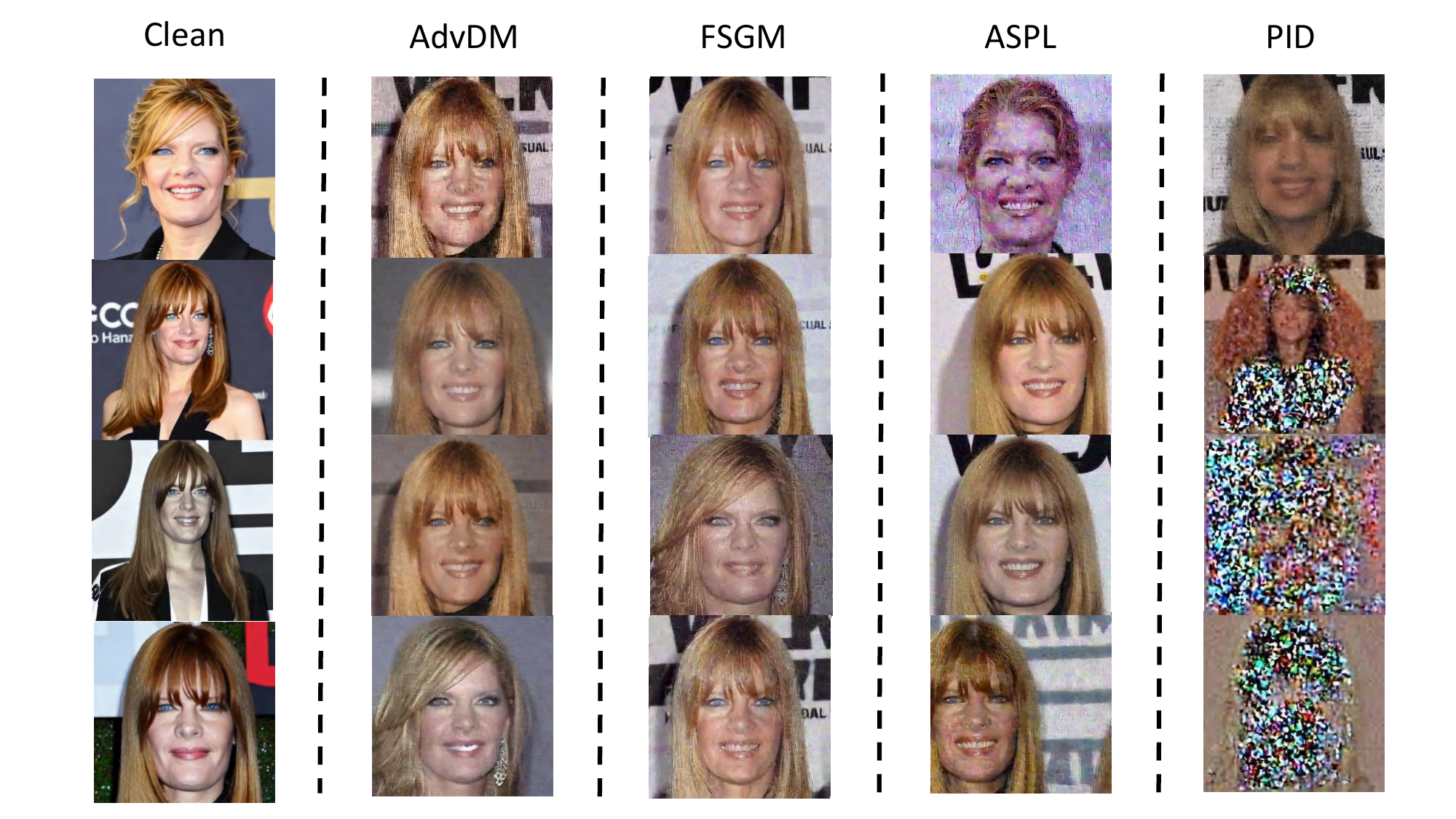}
\end{subfigure}%
\caption{Images synthesized by SD v1.5 fine-tuned on an instance from the CelebA-HQ protected by the defensive algorithms. The text-encoder is trained during fine-tuning and the algorithm is Dreambooth. }
\end{figure}

\begin{figure}[h]
\label{figure 15}
\begin{subfigure}[b]{1.0\textwidth}
    \centering
    \includegraphics[width=0.95\textwidth, page=7]{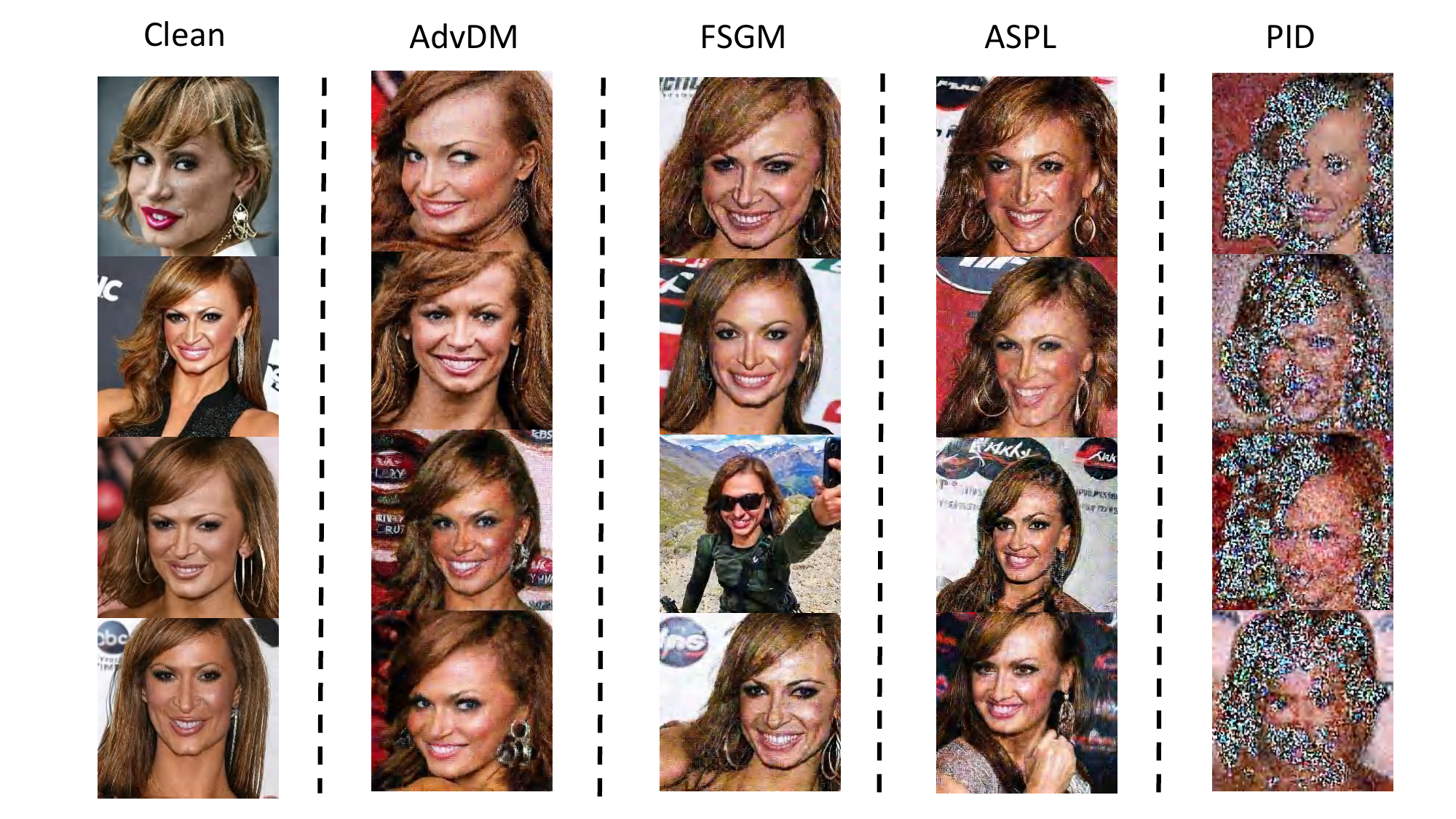}
\end{subfigure}%
\caption{Images synthesized by SD v2.1 fine-tuned on an instance from the CelebA-HQ protected by the defensive algorithms. The text-encoder is frozen during fine-tuning and the algorithm is Dreambooth. }
\end{figure}

\begin{figure}[h]
\label{figure 16}
\begin{subfigure}[b]{1.0\textwidth}
    \centering
    \includegraphics[width=0.95\textwidth, page=8]{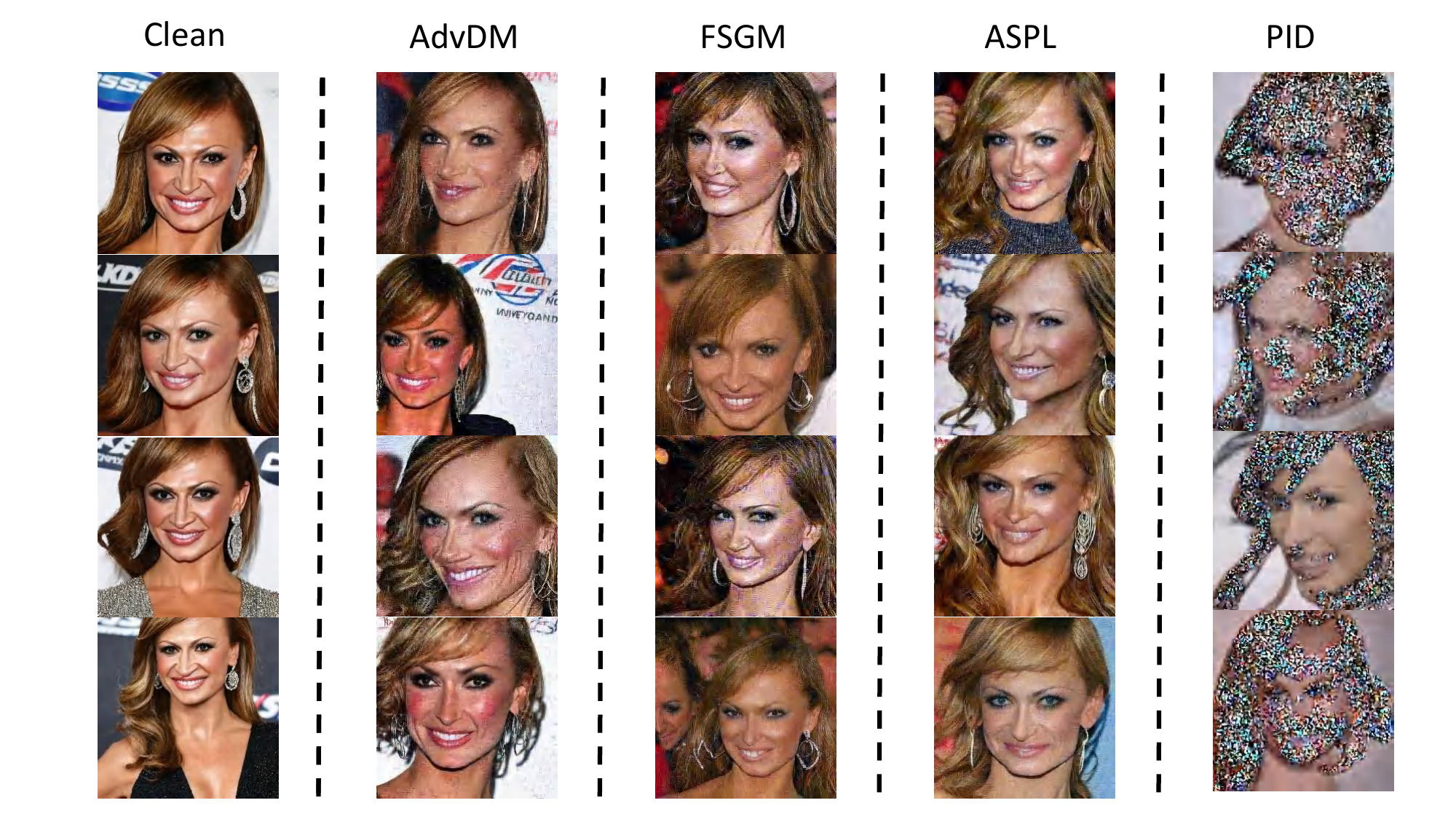}
\end{subfigure}%
\caption{Images synthesized by SD v2.1 fine-tuned on an instance from the CelebA-HQ protected by the defensive algorithms. The text-encoder is trained during fine-tuning and the algorithm is Dreambooth. }
\end{figure}

\begin{figure}[!b]
\label{figure 17}
\begin{subfigure}[b]{1.0\textwidth}
    \centering
    \includegraphics[width=0.95\textwidth, page=9]{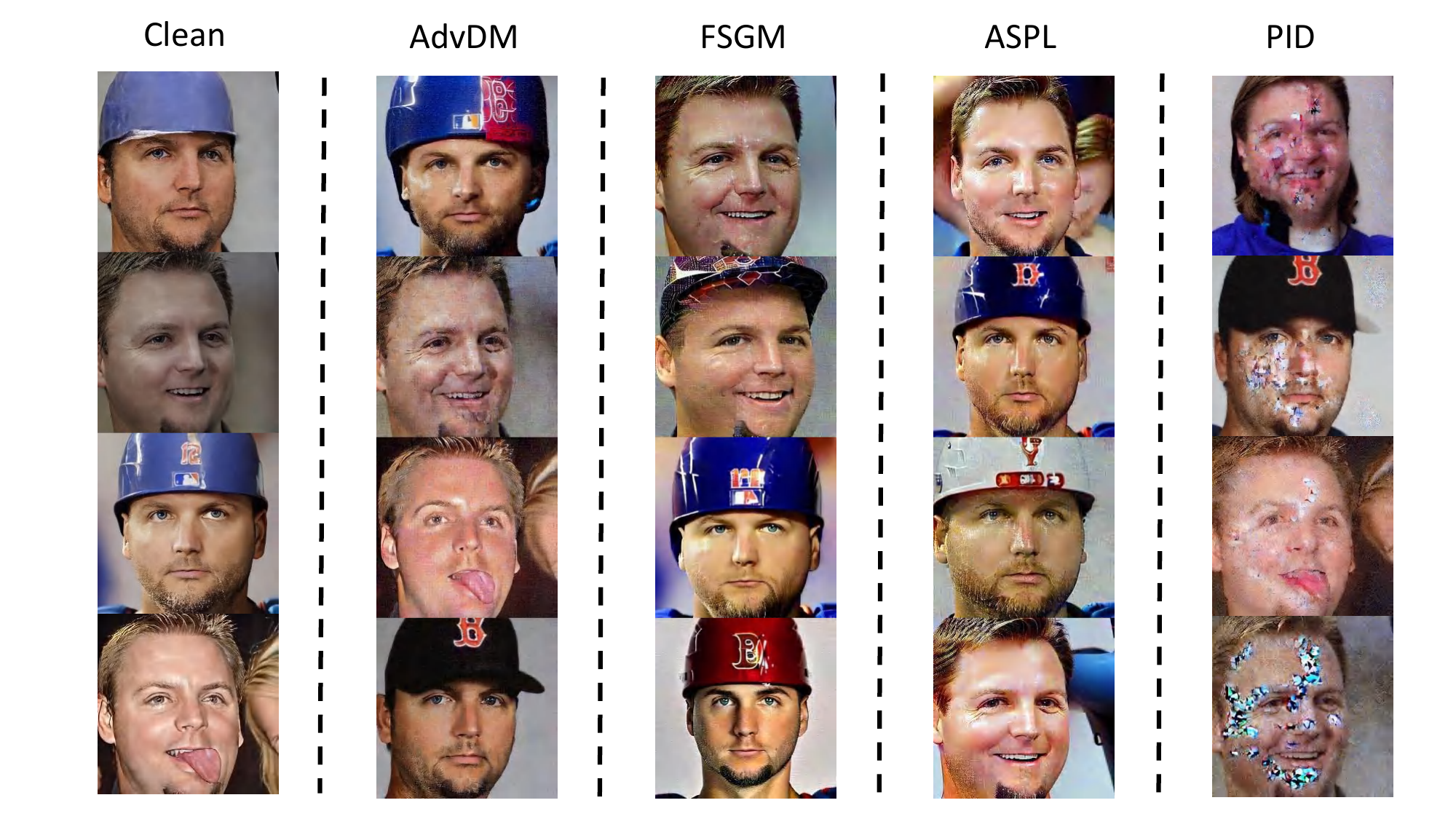}
\end{subfigure}%
\caption{Images synthesized by SD v1.5 fine-tuned on an instance from the VGGFACE protected by the defensive algorithms. The text-encoder is frozen during fine-tuning and the algorithm is Dreambooth. }
\end{figure}

\begin{figure}[h]
\label{figure 18}
\begin{subfigure}[b]{1.0\textwidth}
    \centering
    \includegraphics[width=0.95\textwidth, page=10]{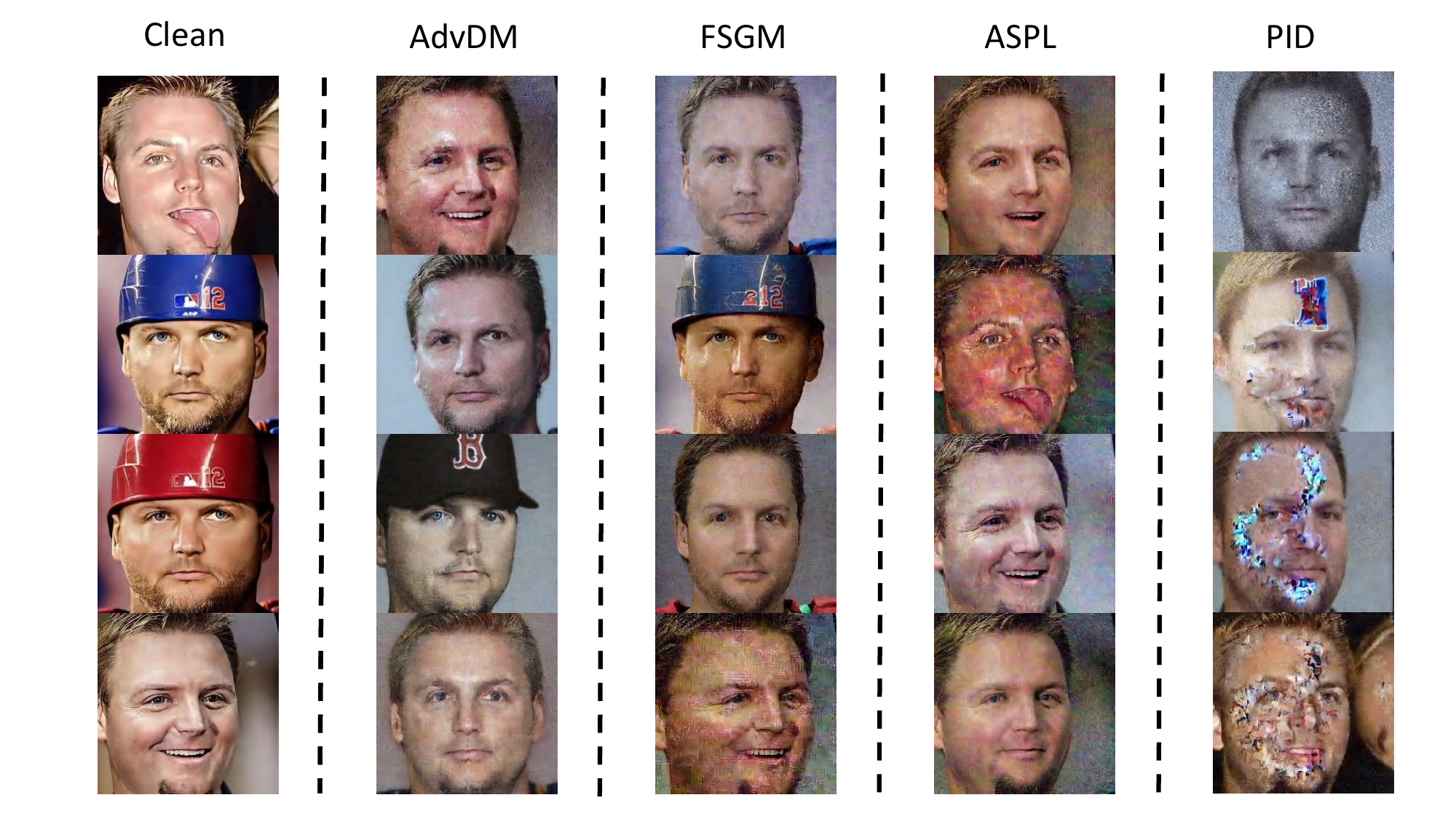}
\end{subfigure}%
\caption{Images synthesized by SD v1.5 fine-tuned on an instance from the VGGFACE protected by the defensive algorithms. The text-encoder is trained during fine-tuning and the algorithm is Dreambooth. }
\end{figure}

\begin{figure}[h]
\label{figure 19}
\begin{subfigure}[b]{1.0\textwidth}
    \centering
    \includegraphics[width=0.95\textwidth, page=11]{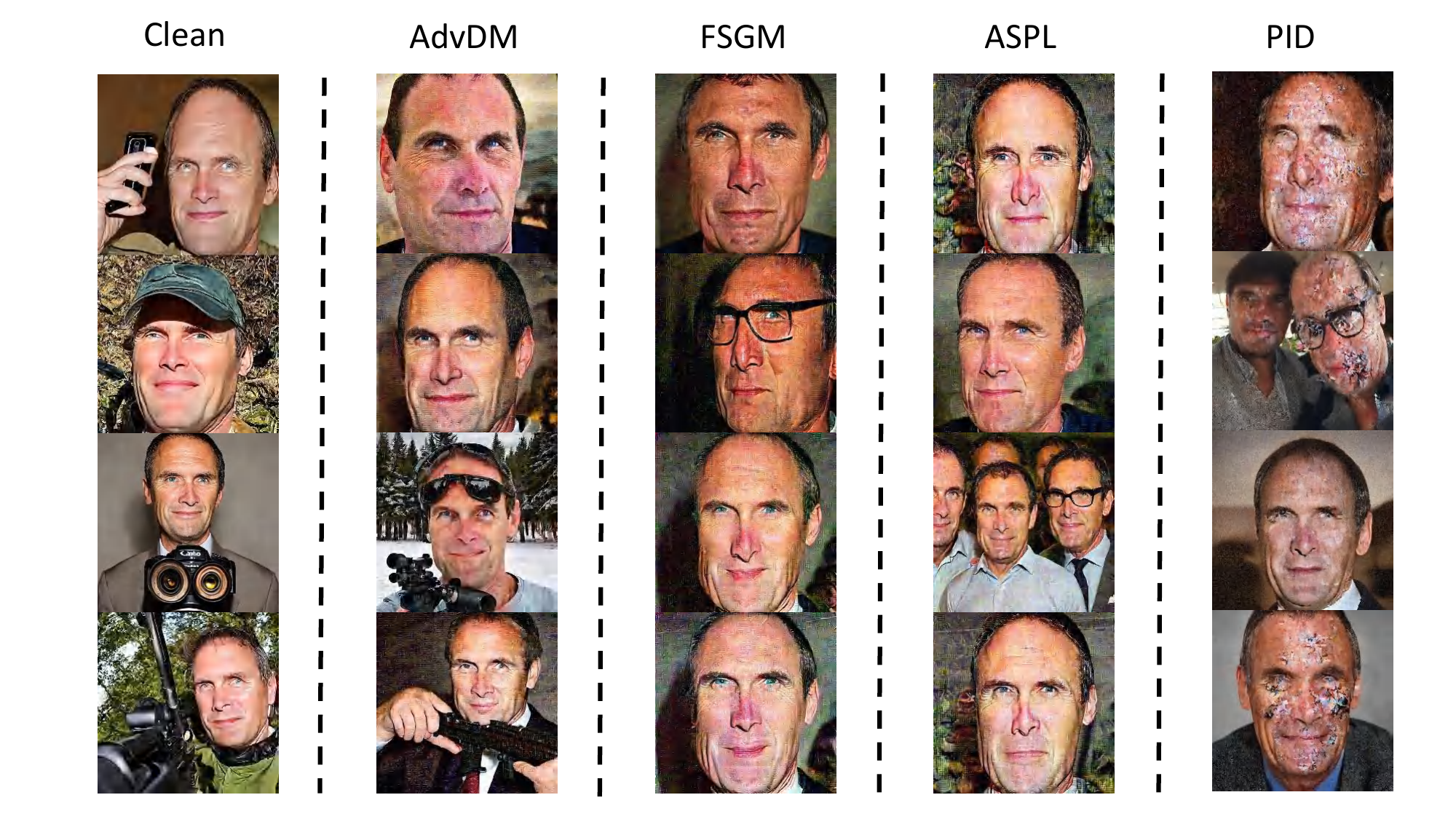}
\end{subfigure}%
\caption{Images synthesized by SD v2.1 fine-tuned on an instance from the VGGFACE protected by the defensive algorithms. The text-encoder is frozen during fine-tuning and the algorithm is Dreambooth. }
\vspace{-1mm}
\end{figure}

\begin{figure}[h]
\label{figure 20}
\begin{subfigure}[b]{1.0\textwidth}
    \centering
    \includegraphics[width=0.95\textwidth, page=12]{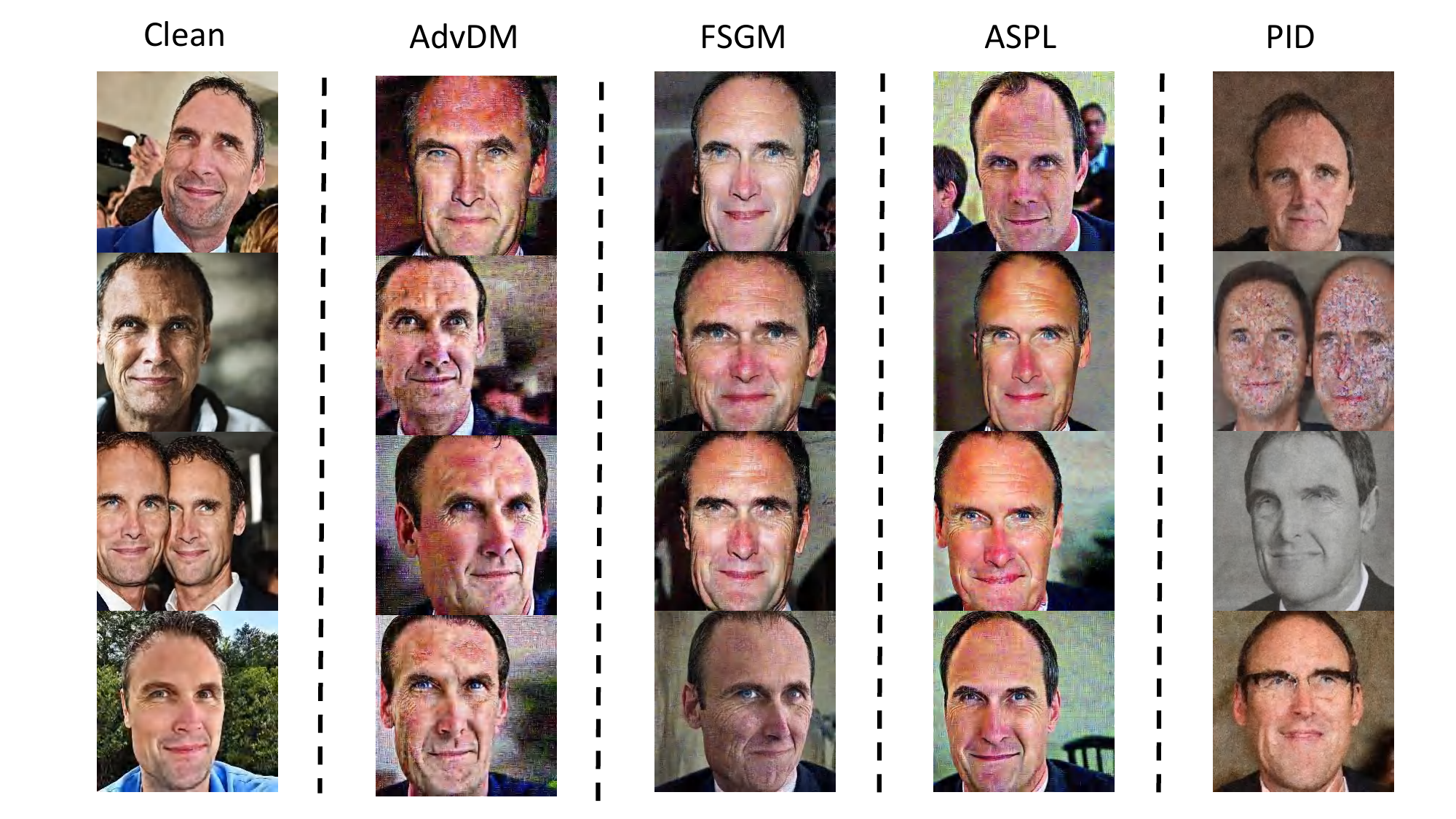}
\end{subfigure}%
\caption{Images synthesized by SD v2.1 fine-tuned on an instance from the VGGFACE protected by the defensive algorithms. The text-encoder is trained during fine-tuning and the algorithm is Dreambooth. }
\end{figure}

\begin{figure}[!b]
\label{figure 21}
\begin{subfigure}[b]{1.0\textwidth}
    \centering
    \includegraphics[width=0.95\textwidth, page=13]{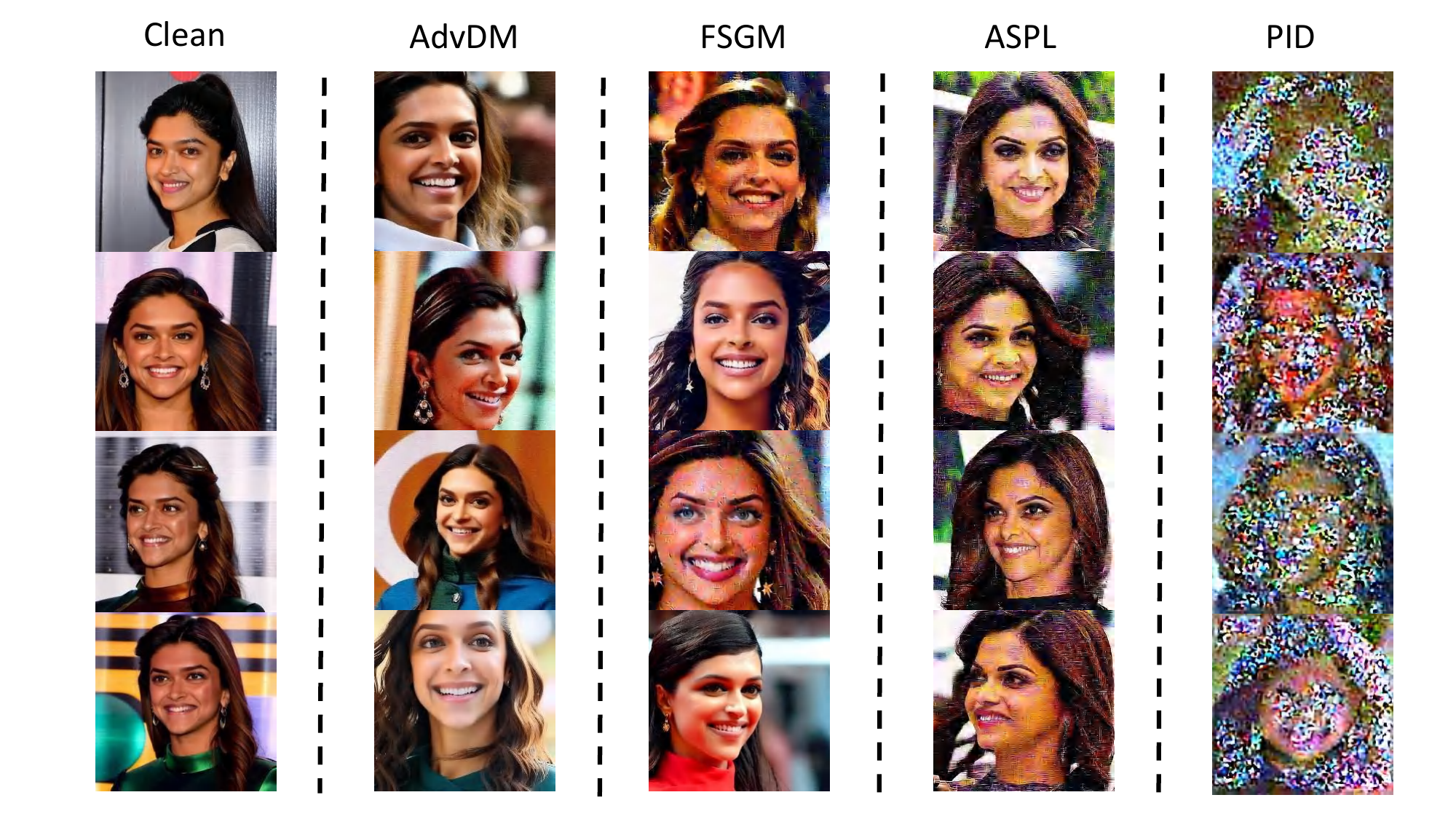}
\end{subfigure}%
\caption{Images synthesized by SD v1.5 fine-tuned on an instance from the CelebA-HQ protected by the defensive algorithms. The fine-tuning algorithm is LoRA. }
\end{figure}

\begin{figure}[h]
\label{figure 22}
\begin{subfigure}[b]{1.0\textwidth}
    \centering
    \includegraphics[width=0.95\textwidth, page=14]{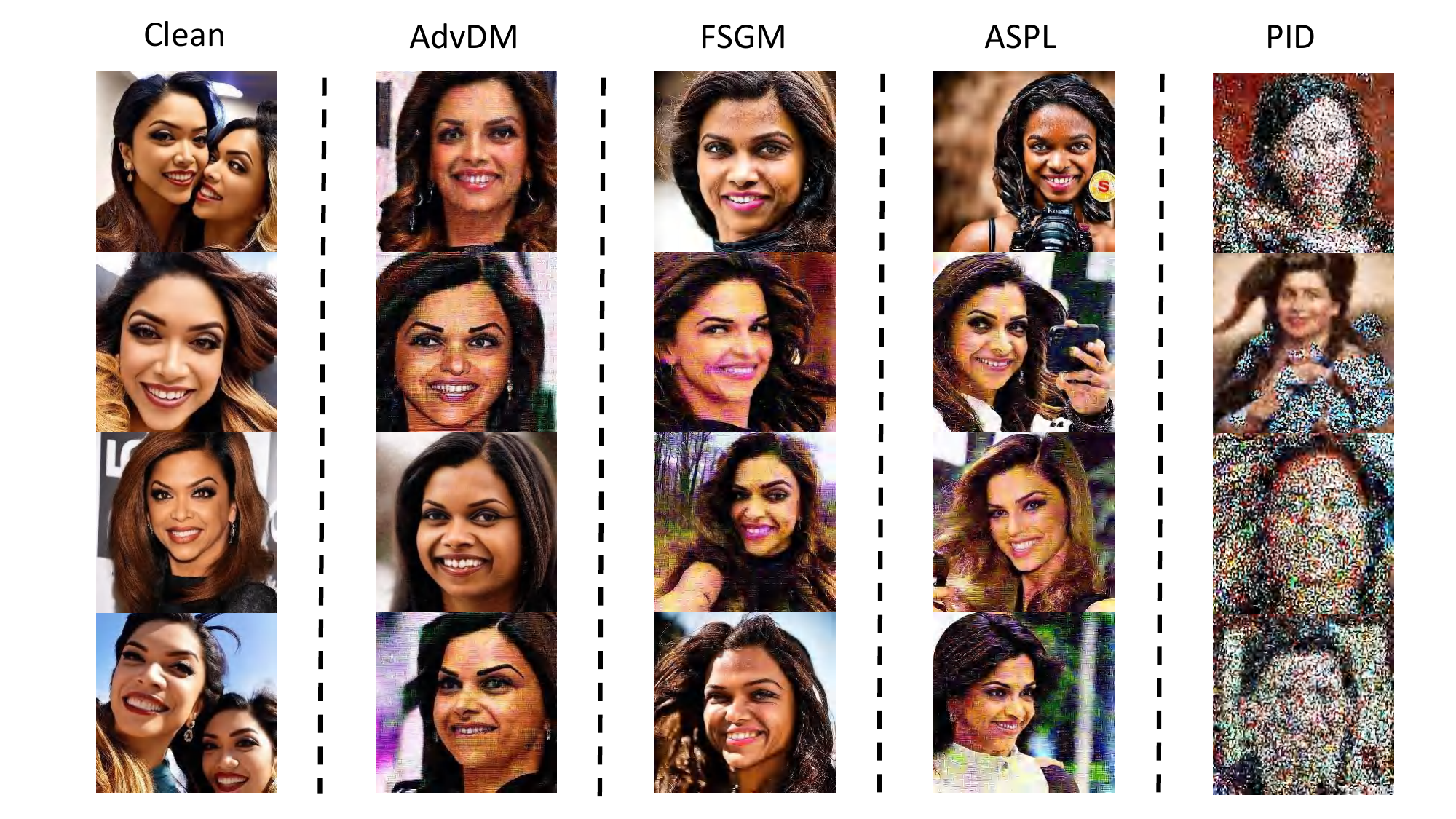}
\end{subfigure}%
\caption{Images synthesized by SD v2.1 fine-tuned on an instance from the CelebA-HQ protected by the defensive algorithms. The fine-tuning algorithm is LoRA. }
\end{figure}

\end{document}

%% file: math_commands.tex
\usepackage{amsmath,amsfonts,bm}

\def\eqref#1{equation~\ref{#1}}

\def\1{\bm{1}}

\def\vx{{\bm{x}}}

\DeclareMathAlphabet{\mathsfit}{\encodingdefault}{\sfdefault}{m}{sl}
\SetMathAlphabet{\mathsfit}{bold}{\encodingdefault}{\sfdefault}{bx}{n}

\def\gE{{\mathcal{E}}}

\def\gN{{\mathcal{N}}}

\newcommand{\E}{\mathbb{E}}

\renewcommand{\epsilon}{\varepsilon}

\newcommand{\bI}{\mathbf{I}}

\newcommand{\norm}[1]{\ensuremath{\left\| #1 \right\|}}